\documentclass[10pt, a4paper]{article}
\usepackage{graphicx}
\usepackage{subfig}
\usepackage{array}
\usepackage{diagbox}
\usepackage{amsmath}
\usepackage{amsfonts}
\usepackage{amssymb}
\usepackage[overload]{empheq}
\DeclareMathOperator{\Diver}{Div}
\DeclareMathOperator{\Grad}{Grad}

\usepackage{mathtools}
\usepackage{pgfplots}
\pgfplotsset{/pgf/number format/use comma,compat=newest}

\newcommand{\R}{\mathbb{R}}

\newcommand{\vect}[1]{\boldsymbol{#1}}
\newcommand{\tens}[1]{\mathsf{#1}}
\begin{document}

\title{Rayleigh--Taylor instability in soft elastic layers}

\author{
D. Riccobelli\textsuperscript{1} and P. Ciarletta\textsuperscript{1},\\
\textsuperscript{1} MOX--Dipartimento di Matematica, Politecnico di Milano, \\
piazza Leonardo da Vinci 32, 20133 Milano, Italy.}



\maketitle

\begin{abstract}
This work investigates the morphological stability of a soft body composed of two heavy elastic layers, attached to a rigid surface and subjected only to the bulk gravity force. Using theoretical and computational tools, we characterize the selection of different patterns as well as their nonlinear evolution, unveiling the interplay between elastic and geometric effects for their formation.

Unlike similar gravity-induced shape transitions in fluids, as the Ray\-leigh--Taylor instability, we prove that the nonlinear elastic effects saturate the dynamic instability of the bifurcated solutions, displaying a rich morphological diagram where both digitations and stable wrinkling can emerge. The results of this work provide important guidelines for the design of novel soft systems with tunable shapes, with several applications in engineering sciences.
\end{abstract}

	\section{Introduction}
	
	Shape transitions in soft solids result from a bifurcation of the elastic solutions driven by either geometrical or constitutive nonlinearities.  The characterization of the emerging morphologies is the object of morpho-elasticity, a recent branch of continuum mechanics at the interface between finite elasticity and perturbation theory. This vibrant research field has rapidly developed in the last decade, pushed by the technological availability of experimental devices controlling the extreme deformations of soft incompressible materials, such as hydrogels \cite{tanaka1987mechanical, dervaux2012mechanical, kim2012designing} and  elastomers \cite{zhu2010large, keplinger2012harnessing}.

Although their boundary value problems are intrinsically different, the study of pattern formation in soft solids has highlighted some similarities, yet several relevant differences, with the instability characteristics of hydrodynamic systems.  For example, if the surface tension in a thin fluid filament triggers the formation of droplets, which spontaneously break down \cite{rayleigh1879capillary}, such a dynamics can be stabilised by elastic effects in soft solid cylinders \cite{mora2010capillarity}, thus driving the emergence of  stable

 \noindent beads-on-a-string patterns \cite{taffetani2015beading}. Similarly, whilst fingering at the interface of two immiscible viscous fluids is an unstable process \cite{saffman1958penetration}, stable digitations may  occur after a subcritical bifurcation for a fluid pushing against an elastic surface \cite{saintyves2013bulk} and at the interface between a thin elastic layer adhering to a glass plate \cite{ghatak2000meniscus}.

Another interesting example is the gravity-induced instability in an elastic layer attached to a rigid substrate with a traction-free surface facing downwards. Contrarily to gravity waves in a fluid layer, the free surface experiences fluctuations that eventually saturate when the large deformations store an elastic free energy of the same order as the corresponding variation of the potential energy. The linear stability analysis of this problem has been recently performed \cite{mora2014gravity}, then refined to consider the effect of an applied strain on the elastic layer \cite{liang2015gravity}. Nonetheless, this problem had been previously solved using numerical techniques \cite{auricchio2013approximation}, often being used as a test case to study  the stability of discrete solutions, obtained by means of mixed finite element techniques \cite{auricchio2005stability, auricchio2010importance}.

 Since Rayleigh \cite{rayleigh83} and Taylor \cite{taylor1950instability}, it is well known that the horizontal interface between one fluid layer put on top of a lighter one is unstable to perturbation of long wavelength, i.e. bigger than the capillary length, forming protrusions growing with a characteristic time. However, if one takes surface tension into account, the growth of small wavelength protrusions is  inhibited by capillary effects, thus larger wavelength drops grow and eventually drip \cite{fermigier1992two}. In this work, we aim at studying this kind of gravity instability in a soft system made of two heavy elastic layers attached on one end to a rigid surface. In particular, we are interested in characterizing both pattern formation and its nonlinear evolution, determining the interplay between elastic and geometric effects for the emergence of a given pattern.

The article is organized as follows. In Section 2, we define the nonlinear elastic problem and identify its basic solution. In Section 3, we perform the linear stability analysis of the problem, deriving the marginal stability curves as a function of the elastic and geometric dimensionless parameters. In Section 4, we perform numerical simulations using finite elements for studying pattern formation in the fully nonlinear regime. In Section 5, we finally discuss the theoretical and numerical results, adding a few concluding remarks. 

	\section{The non-linear elastic problem and its basic solution}
	
\begin{figure}[b!]
	\centering
	\subfloat[$g>0$] {\includegraphics[width=0.5\textwidth]{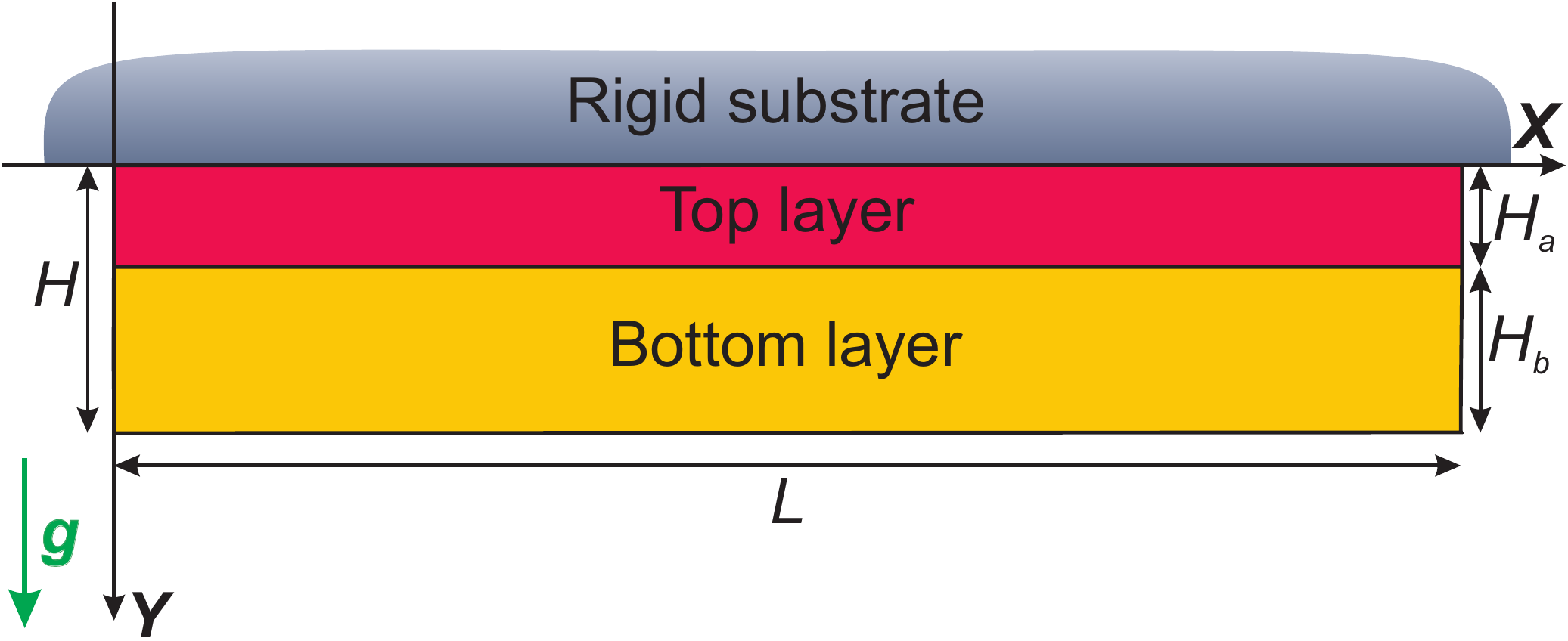}\label{fig:dominio1}}
	\subfloat[$g<0$]{\includegraphics[width=0.5\textwidth]{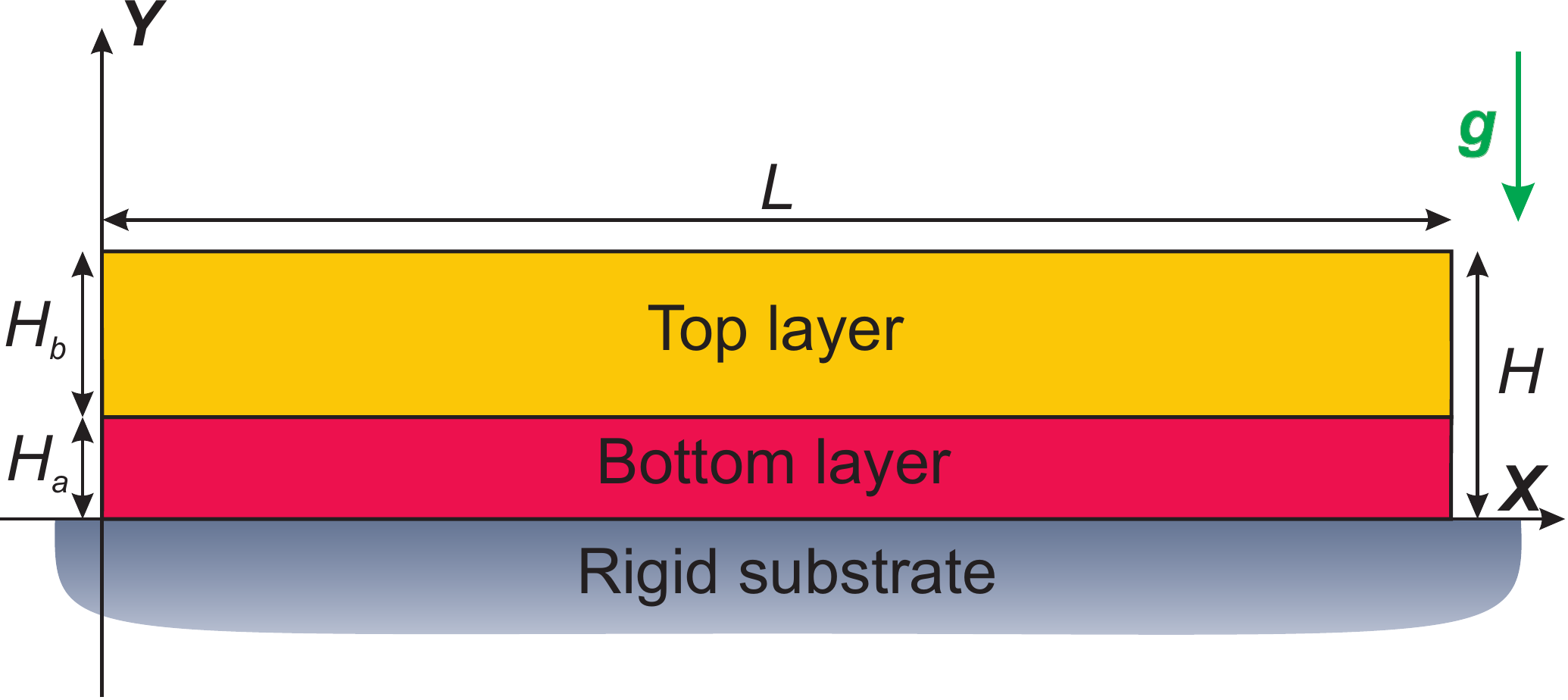}\label{fig:dominio2}}
	\caption{Sketch of the material setting in two different configurations. Case (a):  the body hangs below on a fixed rigid wall, thus being subjected to a tensile gravity force along $Y$ (left). Case (b): the body is placed on top of a rigid substrate, thus being subjected to a compressive gravity force along $Y$ (right).}
	\label{fig:dominio}
	\end{figure}

	In a Cartesian coordinate system with unit base vectors
	$\vect{E}_i$, with {$i=(X,Y,Z)$}, we consider a soft body made of two hyperelastic layers, as sketched in Figure~\ref{fig:dominio}.
	
	Let $\mathbb{E}^3$ be the three-dimensional Euclidean
	space, the body occupies a domain  $\Omega\subset\mathbb{E}^3$, having a thickness $H$ along the Y axis and a length $L$ along the X axis, with $L\gg H$. We also 
consider that the body is infinitely long along the $Z$ direction, so that a plane strain assumption can be made, hence 
\[
\Omega = \left(0,\,L\right)\times\left(0,\,H\right)\times\R.
\]

The  body  is clamped to a rigid substrate at $Y = 0$, 
so that its volume $\Omega$  can be split in the two subdomains $\Omega_a$ and  $\Omega_b$ occupied by the constituting layers, such that:

	\begin{gather*}
	\Omega_a = \left\{\vect{X}\in\Omega\;|\; 0< Y < H_a\right\},\\
	\Omega_b = \left\{\vect{X}\in\Omega\;|\; H_a < Y < H\right\},
	\end{gather*}
where $\vect{X}$ is the material position vector, $H_a$ and $H_b$ are the thicknesses of the layers.
	
	 Indicating
	 $ \vect{x}=\vect{x}(X,Y)$ the spatial position vector,  the kinematics
	 is described by the geometrical deformation tensor $\tens{F}=
	 \Grad \vect{x}$. We also assume that the layers behave as incompressible  neo-Hookean materials and the strain energy density of each layer is given by
	\begin{equation}
	\label{eqn:energia}
	W_\beta(\tens{F})=\frac{\mu_\beta}{2}\left(I_1-2\right) - p(\det\tens{F}-1),\qquad \beta=(a,\,b);
	\end{equation}
	where $I_1$ is the trace of the right Cauchy--Green tensor $\tens{C}=\tens{F}^T\tens{F}$ and $p$ is the Lagrangian multiplier enforcing the internal constraint of incompressibility.	
	
	Using the constitutive assumption in  Eq.~\eqref{eqn:energia}, the first Piola-Kirchhoff stress tensor $\tens{S}$ reads:
	\[
	\tens{S}_\beta= \frac{\partial W_\beta}{\partial \tens{F}}=\mu_\beta \tens{F}^T -p \tens{F}^{-1}, \qquad\beta=(a,\,b).
	\]
	
	Assuming quasi-static conditions, the balance of linear momentum for the elastic body subjected to its own weight reads:
	\begin{equation}
	\label{eqn:fullnonlin}
	\Diver\tens{S}_\beta + \rho_\beta \vect{g}=0\qquad\text{in } \Omega_\beta,\quad\text{where }\beta=(a,\,b);
	\end{equation}
	where $\Diver$ is the material divergence, $\rho_a$ and $\rho_b$ are the densities of the layers, $\vect{g}= g \vect{E}_Y$, is the gravity acceleration vector.

	In the following we aim to provide a unified analysis of the two configurations depicted in Figure~\ref{fig:dominio}. For the sake of notational compactness, we consider that a positive $g$ represents the body hanging down a rigid wall (Figure~\ref{fig:dominio1}), and a negative $g$  the body  placed on top of a rigid substrate (Figure~\ref{fig:dominio2}).
		
	The two boundary conditions at the fixed substrate and at the free surface read
	\begin{equation}
	\label{eqn:bc}
	\left\{
	\begin{aligned}
	&\tens{S}^T_b \vect{E}_Y=\vect{0}\qquad&&\text{for }Y=H,\\
	&\vect{E}_Y\cdot\tens{S}^T_\beta\vect{E}_X=0\qquad&&\text{for }X=(0,\,L), \beta=(a,\,b)\\
	&\vect{u}=\vect{0}\quad&&\text{for }Y=0,\\
	&\vect{u}\cdot\vect{E}_X=0\quad&&\text{for }X=(0,\,L);
	\end{aligned}
	\right.
	\end{equation}
	where $\vect{u}=(\vect{x}-\vect{X})$ is the displacement vector field. The elastic boundary value problem is finally complemented by the following displacement and stress continuity conditions at the interface between the two layers, respectively
	\begin{equation}
	\label{eqn:int}
	\left\{
	\begin{aligned}
	&\lim_{Y\rightarrow H_a^-}\vect{u}=\lim_{Y\rightarrow H_a^+}\vect{u},\\
	&\lim_{Y\rightarrow H_a^-}\tens{S}^T_a\vect{E}_Y=\lim_{Y\rightarrow H_a^+}\tens{S}^T_b\vect{E}_Y.
	\end{aligned}\right.
	\end{equation}
	
	The boundary value problem Eqs.~\eqref{eqn:fullnonlin}--\eqref{eqn:int} admits a basic solution given by
	\begin{equation}
	\label{eqn:basesol}
	\vect{u}=\vect{0},\qquad 
	p = \left\{\begin{aligned}
	&\mu_a + \rho_a g (Y-H_a)-\rho_b g H_b&&\text{for }0< Y < H_a,\\	
	&\mu_b + \rho_b g (Y - H)&&\text{for }H_a < Y < H;
	\end{aligned}
	\right.
	\end{equation}
	so that no basic deformation is allowed by the incompressibility constraint,  and the body is subjected to a hydrostatic pressure linearly dependent on
$Y$. We also highlight that the pressure field in  Eq.~\eqref{eqn:basesol} is discontinuous if $\mu_a\neq\mu_b$ or $\rho_a\neq\rho_b$.
	
	\section{Linear stability analysis of the basic solution}	
	\label{sec:conti}
	\subsection{Incremental equations}
	
	We now aim at investigating the stability of the basic elastic solution Eq.~\eqref{eqn:basesol} using the method of incremental deformations superposed on a finite strain \cite{ogden1997non}.
	
	Let us perturb the basic configuration by applying an incremental displacement $\delta\vect{u}$, if we set $\delta\mathsf{F}=\Grad{\delta\vect{u}}$, the linearised incremental Piola-Kirchhoff stress tensor is
	\[
	\delta\tens{S}_\beta = \mathcal{A}^\beta_0:\delta\tens{F}+p\,\delta\tens{F}-\delta p\, \tens{I}\quad\text{for }\beta=(a,\,b);
	\]
	where
	\[
	\mathcal{A}^\beta_0=\frac{\partial^2 W_\beta}{\partial \tens{F}\partial \tens{F}};\quad \text{ with}\quad\mathcal{A}^\beta_{0ijhk}=\frac{\partial^2 W_\beta}{\partial F_{ji}\partial F_{kh}}
	\]
	is the tensor of instantaneous elastic moduli, $\tens{I}$ is the identity tensor, $\delta p$ is the increment of the Lagrangian multiplier $p$ and the two dots operator ($:$) denotes the double contraction of the indices, namely
	\[
	(\mathcal{A}^\beta_0:\delta\tens{F})_{ij}=\mathcal{A}^\beta_{0ijhk}\delta F_{kh}.
	\]
	
	Recalling that the basic solution is undeformed, the incremental incompressibility and equilibrium equations read, respectively
	\begin{gather}
	\label{eqn:incrstress}
	\Diver \delta\tens{S}_\beta=\vect{0}\quad\text{in }\Omega_\beta,\text{ with }\beta=(a,\,b),\\
	\label{eqn:incompres}
	\Diver \delta\vect{u} = 0\qquad\text{in }\Omega.
	\end{gather}
	
	The incremental counterparts of  two boundary conditions at the fixed substrate and at the free surface may be rewritten as, respectively
    \begin{align}[left=\empheqlbrace]
	\label{eqn:bc1}
	&\delta\tens{S}_b^T\vect{E}_Y=\vect{0}&&\text{for }Y=H,\\
	&\vect{E}_Y\cdot\delta \tens{S}^T_\beta\vect{E}_X=\vect{0}&&\text{for }X=(0,\,L),\,\beta=(a,\,b),\\	
	\label{eqn:bc2}
	&\delta {\vect u}= {\vect 0}&&\text{for }Y=0,\\
	&\delta\vect{u}\cdot\vect{E}_X=0&&\text{for }X=(0,\,L).
 \end{align}
 
	Similarly, the incremental versions of the displacement and stress continuity conditions at the interface read:
	\begin{align}[left=\empheqlbrace]
	\label{eqn:cont12}
	&\lim_{Y\rightarrow H_a^-}\delta\vect{u}=\lim_{Y\rightarrow H_a^+}\delta\vect{u},\\
	\label{eqn:cont3}
	&\lim_{Y\rightarrow H_a^-}\delta\tens{S}_a^T\vect{E}_Y=\lim_{Y\rightarrow H_a^+}\delta\tens{S}_b^T\vect{E}_Y.
	\end{align}
	
	In the following, we derive the solution of the  incremental boundary value problem given by Eqs.~\eqref{eqn:incrstress}-\eqref{eqn:cont3}.
	\subsection{Solution of the incremental boundary value problem}
		
	Let us now assume an ansatz by variable separation in the expression of the incremental displacement, namely 
	\begin{equation}
	\label{eqn:perturb}
	\delta\vect{u}=U(Y)\sin(k X)\vect{E}_X+V(Y)\cos(k X)\vect{E}_Y,
	\end{equation}
	where $k$ is the horizontal spatial wavenumber. We recall that such a functional dependence along the $X$ direction suitably describes both the infinite geometry, for which $k$ is a continuous variable,  and a finite length $L$, so that $k=2 \pi n/L$ with integer mode $n$. 

	From Eq.~\eqref{eqn:incompres} we get that
	\begin{equation}
	\label{eqn:uinfdiv}
	k U(Y) = -V'(Y).
	\end{equation}
	
	From the first component of Eq.~\eqref{eqn:incrstress} we obtain the expression for $\delta p$ as
	\begin{equation}
	\label{eqn:incrpres}
	\delta p = \cos(k X)\left(\rho_\beta g V(Y) -\mu_\beta V'(Y)+k^{-2}\mu_\beta V'''(Y)\right) \quad\text{in }\Omega_\beta\text{ with }\beta=(a,\,b).
	\end{equation}
	
	By substituting Eqs.~\eqref{eqn:uinfdiv} and \eqref{eqn:incrpres} in the second component of Eq.~\eqref{eqn:incrstress}, we obtain the following ordinary differential equation:
	\begin{equation}
	\label{eqn:V}
	V''''(Y)-2 k^2 V''(Y)+k^4 V(Y)=0,
	\end{equation}
	which is valid for both layers and whose solution is given by:
	\begin{equation}
	V(Y) = C_{1\beta}e^{-kY}+C_{2\beta}Ye^{-kY}+C_{3\beta}e^{kY}+C_{4\beta}Ye^{kY}\quad\text{in }\Omega_\beta,\,\beta=(a,\,b).
	\end{equation}
	
	 Hence, setting
	 \[
	 \vect{v}=\left[\frac{C_{1a}}{H_a},\,C_{2a},\,\frac{C_{3a}}{H_a},\,C_{4a},\,\frac{C_{1b}}{H_a},\,C_{2b},\,e^{2 k H_a}\frac{C_{3b}}{H_a},\,e^{2 k H_a}C_{4b}\right]^T
	 \]
	 we impose the conditions given in Eqs.~\eqref{eqn:bc1}--\eqref{eqn:cont3}, we find $8$ linear algebraic equations in the unknowns $v_j$, $j=(1,\dots,8)$, so that we can write such system in the compact form $\tens{M}\vect{v}=\vect{0}$ where $\tens{M}$ is the $8\times 8$  coefficients' matrix. Hence, we find that a non-null solution of such linear system exists if and only if 
	\begin{equation}
	\label{eqn:matrice}
	\det \tens{M} = 0;
	\end{equation}
	The full form of $\tens{M}$ is reported in the Appendix \ref{sec:AA}.
	
	\subsection{Results of the linear stability analysis}	
	\label{sec:LSA}

	Let us now discuss the results of the linear stability analysis by making use of the following dimensionless parameters:
	\[
		\gamma = \frac{\rho_a g H_a}{\mu_a},\qquad \alpha_H =\frac{H_b}{H_a},\qquad \alpha_\mu=\frac{\mu_b}{\mu_a},\qquad\alpha_\rho=\frac{\rho_b}{\rho_a},\qquad \tilde{k}=H_a k.
	\]	

\begin{figure}[b!]
	\centering
	\subfloat[$\alpha_\mu=0.5$] {\includegraphics[width=0.49\textwidth]{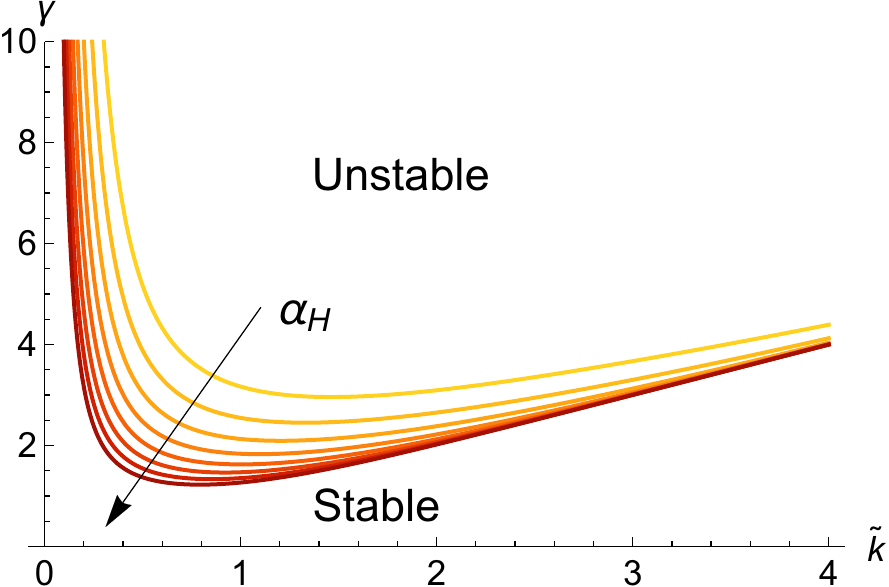}\label{fig:am_0.5}}
	\subfloat[$\alpha_H=1$]{\includegraphics[width=0.49\textwidth]{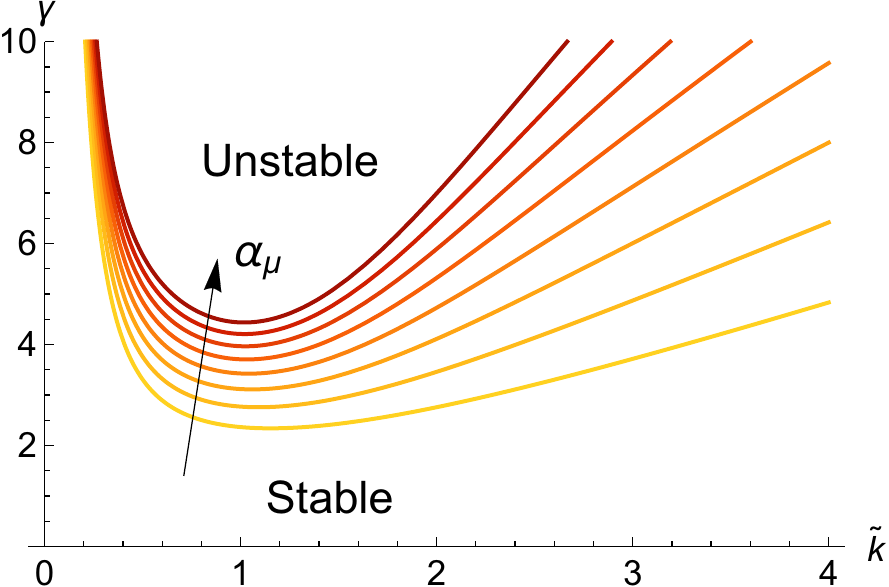}\label{fig:am_2}}\\
	\caption{Marginal stability curves showing the order parameter $\gamma$ versus the horizontal wavenumber $\tilde{k}$ for $\rho_a=\rho_b$ and: (a) $\alpha_\mu=0.5$, (b) $\alpha_H=1$.  The curves are shown at varying  $\alpha_H$ (a) and $\alpha_\mu$ (b) from $0.6$ to $2$ by steps of $0.2$, the arrow indicates the direction in which the parameter grows.}
	\label{fig:ar_1_am}
	\end{figure}

	A great simplification arises if we set both $\alpha_\rho=1$ and $\alpha_\mu=1$ or if we impose $\alpha_H=0$ in Eq.~\eqref{eqn:matrice}, so that the body is made of a single homogeneous slab. In particular, we recover the same expression reported in \cite{mora2014gravity}:
	\[
	\frac{\rho g H}{\mu_a}  = \frac{2k H \left(2 \left(k H\right)^2+\cosh (2 k H)+1\right)}{\sinh (2 k H)-2 k H};
	\]
	highlighting that an elastic bifurcation occurs for the critical value $\frac{\rho g H}{\mu_a}\simeq6.22$ with critical wavenumber $k H\simeq2.11$.
	
		Let us now analyse the resulting solutions when $\alpha_\rho=1$, namely assuming that the body force is the same for both layers. In the case in which $\alpha_\mu\neq1$ or $\alpha_H\neq1$, we find only one root of equation Eq.~\eqref{eqn:matrice}. In Figure~\ref{fig:ar_1_am}  we depict the resulting marginal stability curves  varying the parameters $\alpha_\mu$ and $\alpha_H$.

	\begin{figure}[t!]
	\centering
	\subfloat[$\gamma_{cr}$ vs $\alpha_\mu$] {\includegraphics[width=0.49\textwidth]{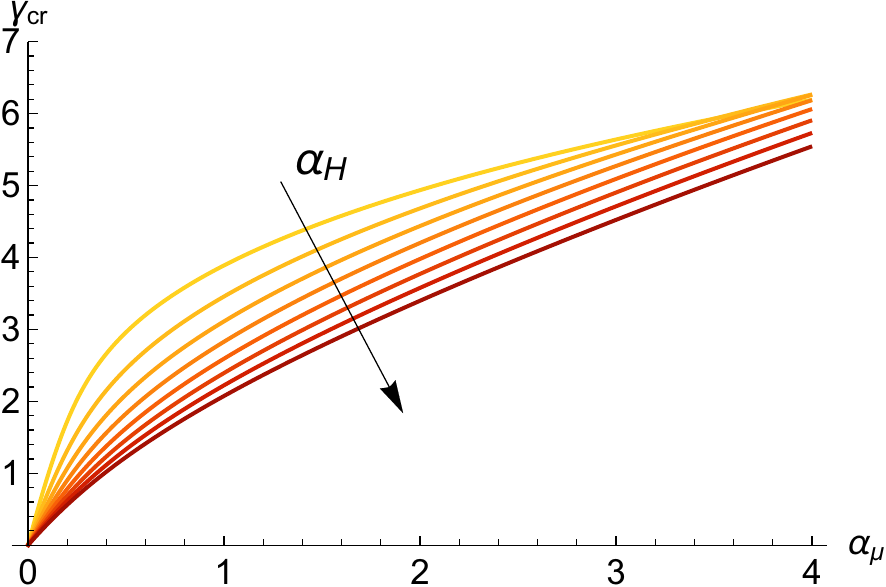}\label{fig:ah_cr}}
	\subfloat[$k_{cr}$ vs $\alpha_\mu$] {\includegraphics[width=0.49\textwidth]{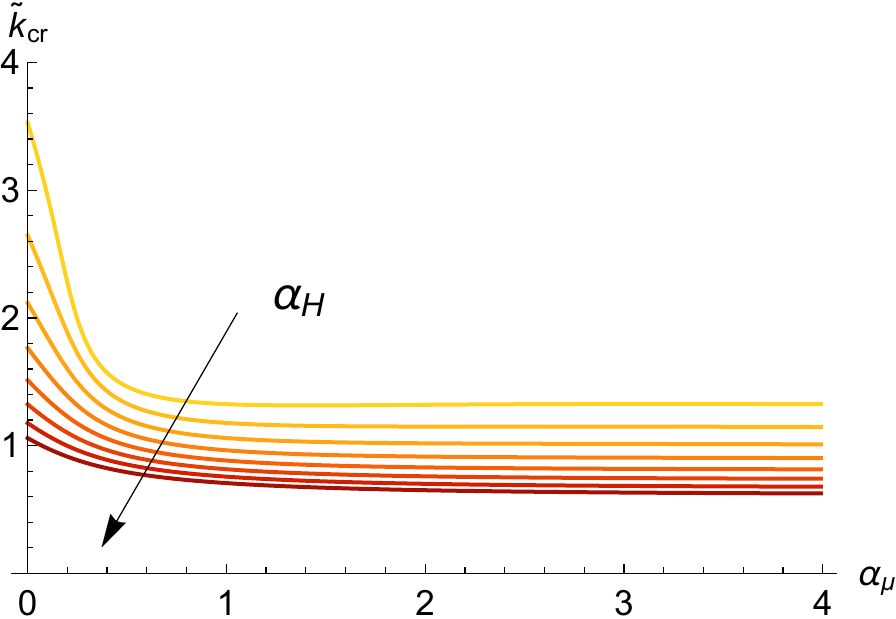}\label{fig:ah_k_cr}}\\
	\subfloat[$\gamma_{cr}$ vs $\alpha_H$] {\includegraphics[width=0.49\textwidth]{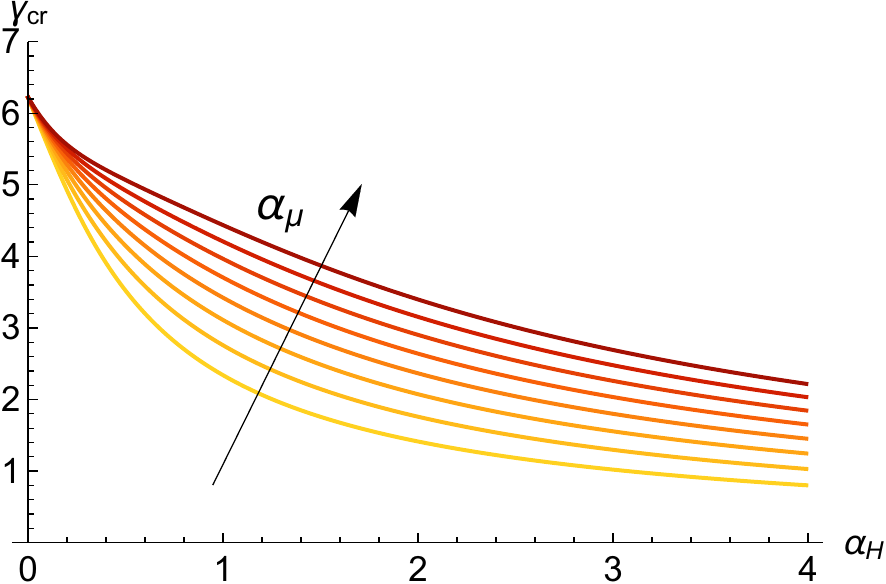}\label{fig:am_cr}}
	\subfloat[$k_{cr}$ vs $\alpha_H$] {\includegraphics[width=0.49\textwidth]{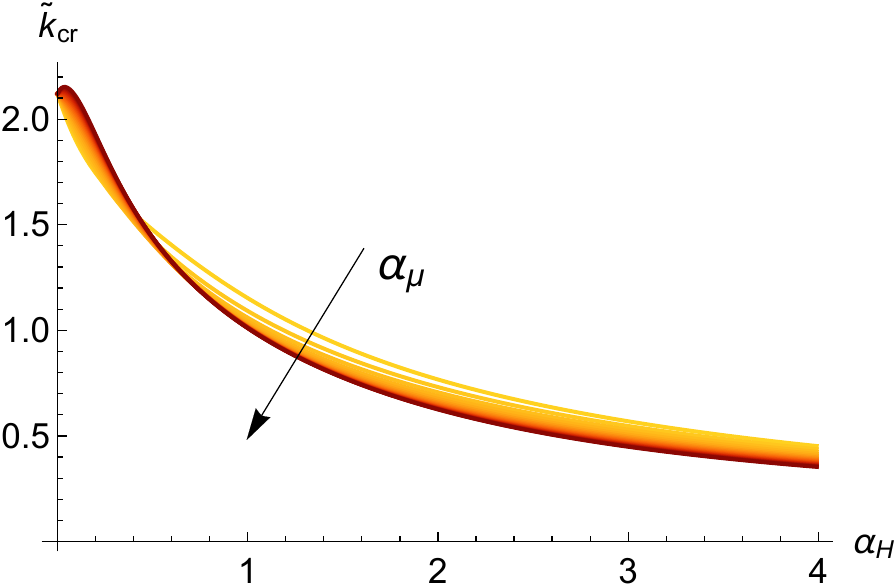}\label{fig:am_k_cr}}
	\caption{Plot of the critical values of (a) $\gamma_{cr}$ versus $\alpha_\mu$ at varying $\alpha_H$; (b) $\gamma_{cr}$ versus $\alpha_H$ at varying $\alpha_\mu$; (c) $\tilde{k}_{cr}$ versus $\alpha_\mu$ at varying $\alpha_H$; (d) $\tilde{k}_{cr}$ versus $\alpha_H$ at varying  $\alpha_\mu$. The arrows indicate the direction in which the parameters $\alpha_H$ (a,b) and $\alpha_\mu$ (c,d) grow from $0.6$ to $2$ by steps of $0.2$.}
	\label{fig:critical}
	\end{figure}	
	
We denote by $\gamma_{cr}$ the critical value of $\gamma$, i.e. the minimum value of the marginal stability curve obtained fixing $\alpha_H$ and $\alpha_\mu$. We denote by $\tilde{k}_{cr}$ the critical wavenumber, namely the value of $\tilde{k}$ for which the marginal stability curve has a minimum. All the critical values of the marginal stability curves have been found by using the Newton's method with the software
\emph{Mathematica} 11.0 (Wolfram Research,Champaign,IL, USA).		
	
	In Figure~\ref{fig:critical} we plot the critical values $\gamma_{cr}$ and $\tilde{k}_{cr}$ when varying $\alpha_H$ and $\alpha_\mu$. We find that $\gamma_{cr}$ strongly depends on $\alpha_\mu$ and $\alpha_H$. In Figure~\ref{fig:ah_cr} we find that if we increase the parameter $\alpha_\mu$ the critical value $\gamma_{cr}$ also increases, so that high values of $\alpha_\mu$ have a stabilizing effect. On the contrary,  in Figure~\ref{fig:am_cr} we find that if we increase $\alpha_H$ the critical value $\gamma_{cr}$ decreases. We highlight that, if $\alpha_H$ tends to zero we obtain that $\gamma_{cr}\simeq 6.22$, which is the single layer limit discussed before. The same limit is found for  $\alpha_\mu$ tending to zero, since it represents the case in which the bottom layer in Figure~\ref{fig:dominio1} becomes infinitely soft. The critical wavelength is always of the same order of the body thickness,  resulting to be more influenced by the parameter $\alpha_H$ if $\alpha_\mu>1$, as we can notice from Figures \ref{fig:ah_k_cr} and \ref{fig:am_k_cr}.
	
	The case in which $\alpha_\rho=1$ is of particular interest in the applications because it is reproducible in experiments using hydrogels. In fact, these soft materials are mainly composed of water, thus having a density which is of the order of $10^3$ kg/m$^3$. Nonetheless, by small variation of the crosslink concentration, it is possible to obtain a shear modulus $\mu$ ranging from $100$ Pa to $10$~kPa. For example, if we consider two  hydrogel layers with $H_a=H_b$ (Figure~\ref{fig:am_2}), where the clamped one has $\mu_a=300$~Pa and the other  $\mu_b=600$~Pa, we find that $\gamma_{cr}\simeq 4.4366$. Accordingly, an instability would appear at $H_a\geq\mu_a \gamma_{cr}/(\rho g)\simeq 13.57\,$cm.

	The general case in which $\alpha_\rho\neq1$ is a bit more complex, in fact Eq.~\eqref{eqn:matrice} can be written in the following compact form
	\begin{equation}
	\label{eqn:eqdi2grado}
	c_1 \gamma^2 + c_2 \gamma + c_3 = 0
	\end{equation}
	where the coefficients $c_1$, $c_2$ and $c_3$ depend on $\alpha_H$, $\alpha_\mu$, $\alpha_\rho$ and $\tilde{k}$, as reported in the Appendix \ref{sec:AB}.
	
	Even if their expressions are very cumbersome, we can still make some general observations. In fact we observe that $c_1$ does not depend on $\alpha_\rho$ whereas, if we fix the other variables, $c_3$ has a different sign if $\alpha_\rho>1$ or if $0<\alpha_\rho<1$. Hence,  one of the two real roots of Eq.~\eqref{eqn:eqdi2grado} changes sign if we consider $\alpha_\rho>1$ or $0<\alpha_\rho<1$.
	
	Thus, we make a distinction in the following between these two cases, which physically correspond to the two configurations depicted in Figure~\ref{fig:dominio}.
	
	\subsubsection{Case (a): free surface instability ($\gamma>0$)}
	
	The configuration shown in Figure~\ref{fig:dominio1} undergoes a morphological transition if Eq.~\eqref{eqn:eqdi2grado} possesses at least a positive root for $\gamma$, since we assume $g>0$.		
	
\begin{figure}[b!]
	\centering
	\subfloat[$\alpha_H=1$, $\alpha_\rho=0.5$] {\includegraphics[width=0.49\textwidth]{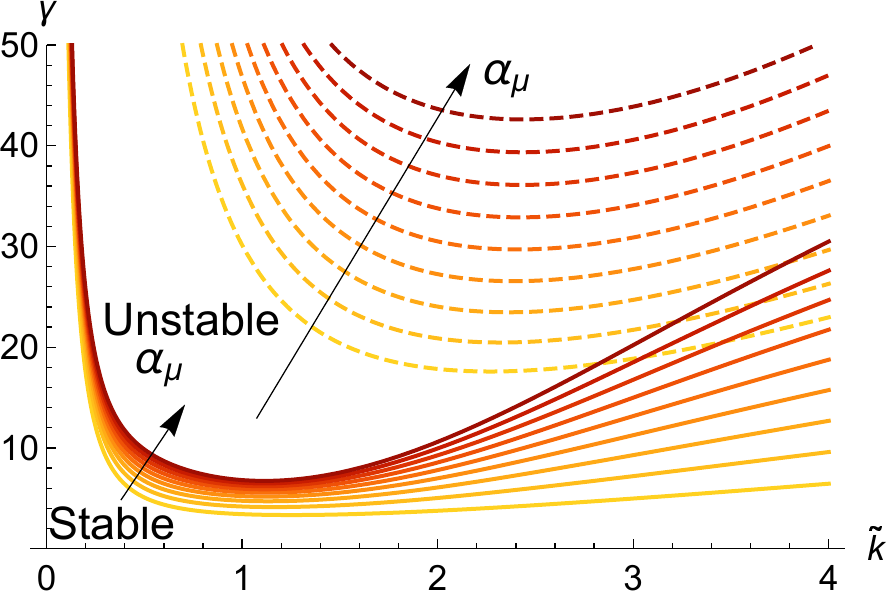}\label{fig:ar_0.5_ah_1}}
	\subfloat[$\alpha_H=1$, $\alpha_\rho=2$]{\includegraphics[width=0.49\textwidth]{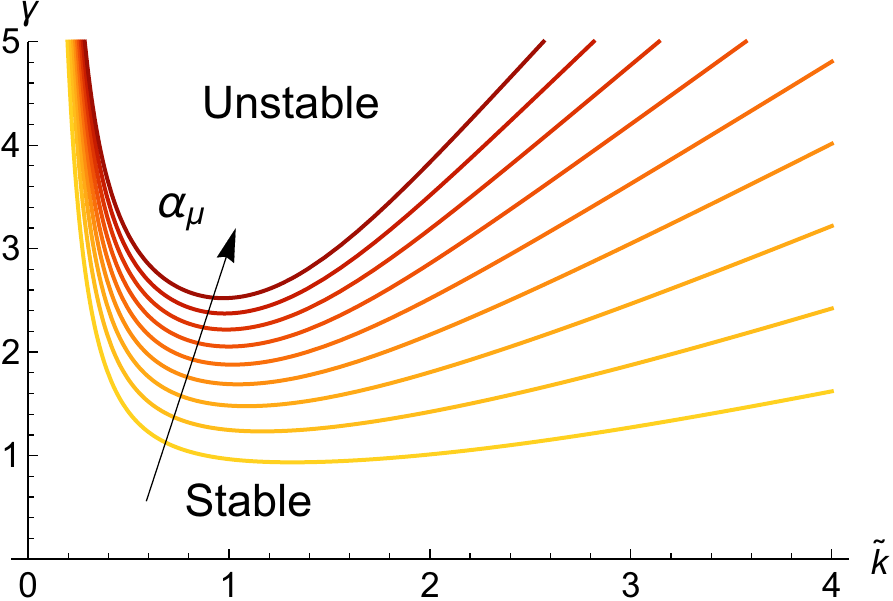}\label{fig:ar_2_ah_1}}
	\caption{Marginal stability curves showing the order parameter $\gamma$ versus the horizontal wavenumber $\tilde{k}$ for $\alpha_H=1$ and: (a) $\alpha_\rho=0.5$, (b) $\alpha_\rho=2$ where $\alpha_\mu$ varies from $0.4$ to $2$ by steps of $0.2$. In (a) we find two positive solutions (solid and dashed lines) of equation Eq.~\eqref{eqn:eqdi2grado} whereas in Figure~(b) we  only find one positive solution.}
	\label{fig:ah_1}
	\end{figure}	
	\begin{figure}[b!]
	\centering
	\subfloat[$\alpha_\mu=1$, $\alpha_\rho=0.5$] {\includegraphics[width=0.49\textwidth]{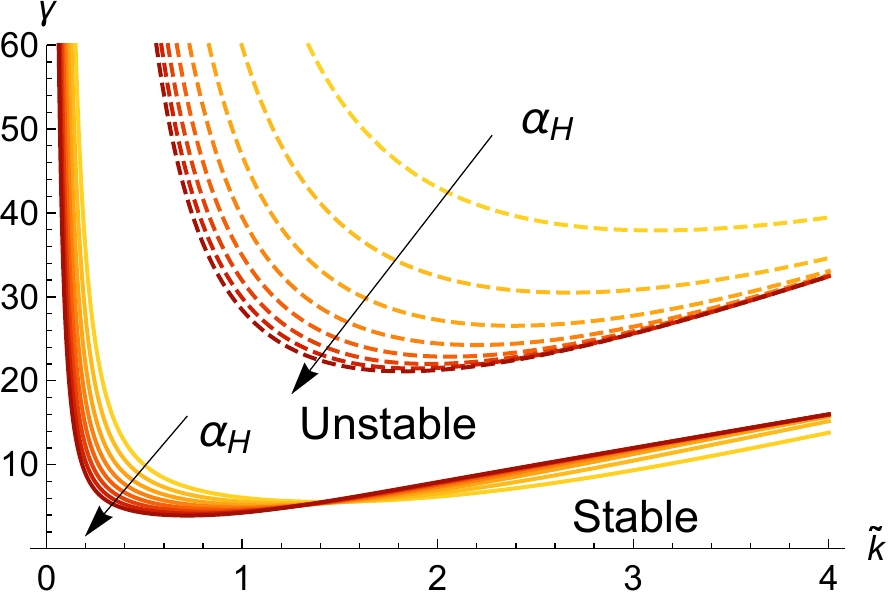}\label{fig:ar_0.5_am_1}}
	\subfloat[$\alpha_\mu=1$, $\alpha_\rho=2$]{\includegraphics[width=0.49\textwidth]{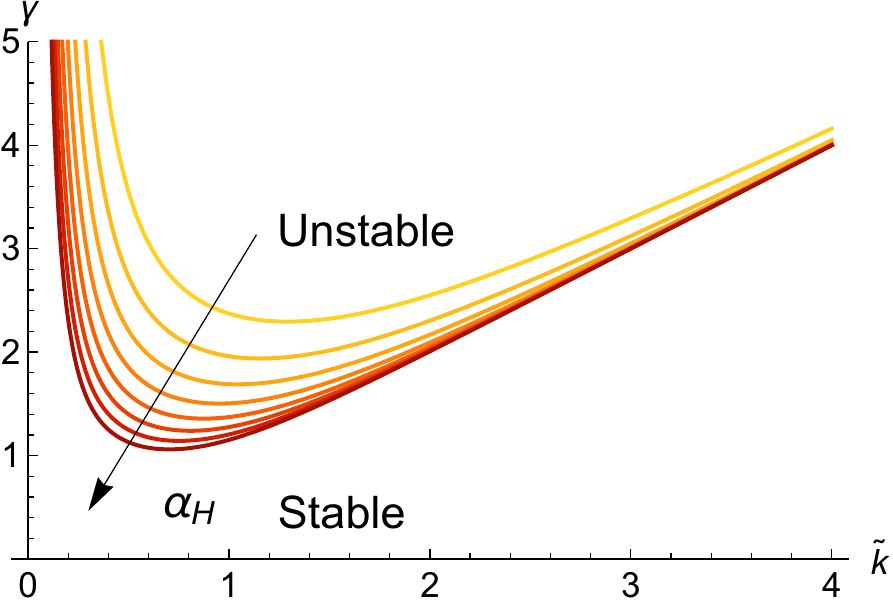}\label{fig:ar_2_am_1}}
	\caption{Marginal stability curves showing the order parameter $\gamma$ versus the horizontal wavenumber $\tilde{k}$ for $\alpha_\mu=1$ and: (a) $\alpha_\rho=0.5$, (b) $\alpha_\rho=2$ with $\alpha_H$ varying from $0.6$ to $2$ by steps of $0.2$. In (a) we find two positive solutions (solid and dashed lines) of equation Eq.~\eqref{eqn:eqdi2grado} whereas in Figure~(b) we  only find one positive solution.}
	\label{fig:am_1}
	\end{figure}

	In Figures \ref{fig:ah_1} and \ref{fig:am_1} we depict the marginal stability curves $\gamma(\tilde{k})$ when we vary the parameters $\alpha_H$, $\alpha_\rho$ and $\alpha_\mu$ in Eq.~\eqref{eqn:eqdi2grado}.  In this case, we find that the instability is localised at the free boundary of the slab, i.e. at $Y=H$.
	
	We can observe that we have the same behaviour discussed for the case $\alpha_\rho=1$: if we increase the parameter $\alpha_\mu$ we obtain a stabilization of the system (i.e. $\gamma_{cr}$ increases) whereas if we decrease $\alpha_H$ we have instability for lower values of $\gamma$.
	
	\subsubsection{Case (b): interfacial instability ($\gamma<0$)}
	
	\begin{figure}[t!]
	\centering
	\subfloat[$\alpha_H=1$, $\alpha_\rho=2$] {\includegraphics[width=0.49\textwidth]{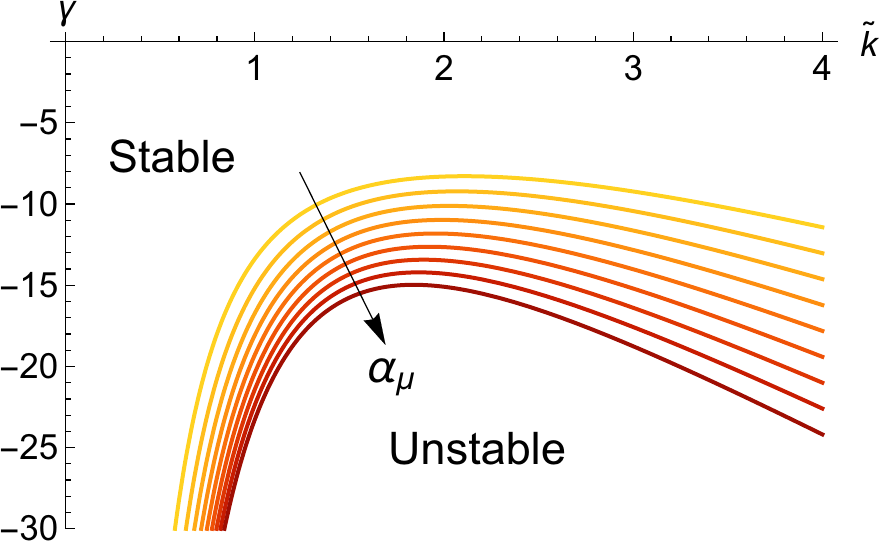}\label{fig:ar_2_ah_1_neg}}
	\subfloat[$\alpha_\mu=1$, $\alpha_\rho=2$]{\includegraphics[width=0.49\textwidth]{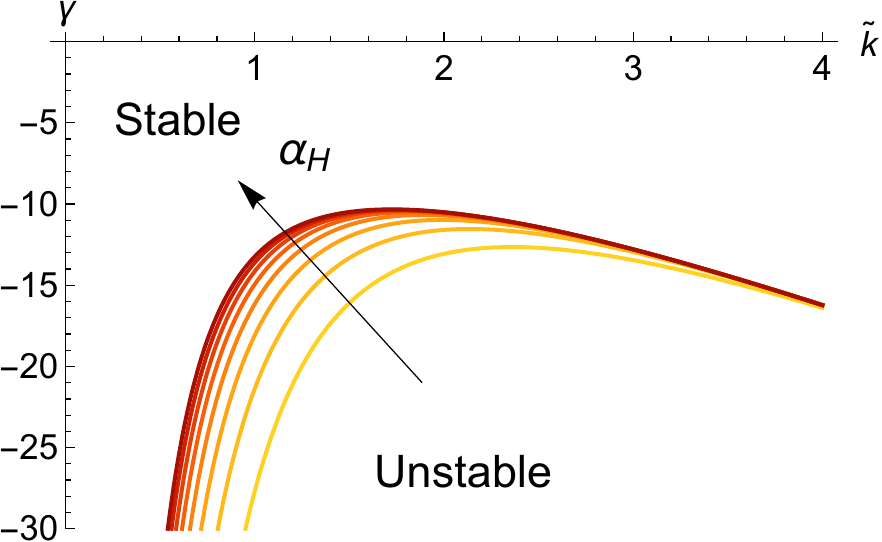}\label{fig:ar_2_am_1_neg}}
	\caption{Marginal stability curves showing the negative root of the order parameter $\gamma$ versus the horizontal wavenumber $\tilde{k}$ for $\alpha_\rho=2$ and: (a) $\alpha_H=1$ with $\alpha_\mu$ varying from $0.4$ to $2$ by steps of $0.2$; (b) $\alpha_\mu=1$ with $\alpha_H$ varying from $0.6$ to $2$ by steps of $0.2$.}
	\label{fig:ar_2_neg}
	\end{figure}	
	
	Conversely, the configuration shown in Figure~\ref{fig:dominio2} undergoes a morphological transition if Eq.~\eqref{eqn:eqdi2grado} possesses a negative root for $\gamma$, since we assume $g<0$. As previously discussed, this happens only if $\alpha_\rho>1$, meaning that the top layer is heavier than the bottom one. Thus, this case is the elastic analog of the Rayleigh-Taylor instability. As found in fluids, the instability is concentrated at the interface between the layers and decays away from it.
	
	In this configuration, we define the critical value $\gamma_{cr}$ as the maximum of the marginal stability curve $\gamma(\tilde{k})$ for fixed $\alpha_H,\,\alpha_\mu,$ and $\alpha_\rho$.
	
	 In Figure~\ref{fig:ar_2_neg} we set $\alpha_\rho=2$ and we plot the marginal stability curves for several values of the parameters $\alpha_H$ and $\alpha_\mu$. Also in this case we highlight that increasing the parameter $\alpha_\mu$ stabilizes the system, whereas an increase of the parameter $\alpha_H$ favours the onset of the interfacial instability.
	
	In the next section, based on the results of the linear stability analysis, we build the simulation tools for studying the fully nonlinear morphological transition.
	
	\section{Post-buckling analysis}
	
	In this section we numerically implement the fully non-linear problem given by Eqs.~\eqref{eqn:fullnonlin}-\eqref{eqn:int}. We finally report the results of 
	numerical simulations for the two cases under considerations, highlighting the morphological evolution of the emerging patterns in the fully nonlinear regime.
	
	\subsection{Finite element implementation}
	
The boundary value problem is implemented by using the open source tool for solving partial differential equations FEniCS \cite{logg2012automated}. In order to enforce the incompressibility constraint, a mixed formulation has been chosen. 
If the two layers have different stiffness or mass density, the pressure field may present a discontinuity at the interface between the two layers, according to the basic solution Eq.~\eqref{eqn:basesol}. Accordingly, we used the element $P_2$--$P_0$ \cite{boffi2013mixed} in numerical simulations. 
	
	This element discretizes the displacement with piecewise quadratic functions and the pressure field with piecewise constant functions, so that we can correctly account for a discontinuous pressure field. It is also numerically stable in linear elasticity  \cite{boffi2013mixed} and it has been successfully used in several non-linear applications \cite{auricchio2005stability, auricchio2013approximation}.
	
	We use a rectangular mesh whose height is $H=1$ and whose length is the critical wavelength $\lambda=2 \pi H_a/\tilde{k}_{cr}$, where $\tilde{k}_{cr}$ is the critical value arising from the previous linear stability analysis and depending on $\alpha_\rho$, $\alpha_H$ and $\alpha_\mu$. We set $\vect{u}=\vect{0}$ at $Y=0$ and we impose periodic boundary conditions at $X=0$ and $X=\lambda$. The number of elements used depends on the length of the mesh, the maximum number of elements used is 30000.
	
	In order to investigate the post-buckling regime, we impose a sinusoidal imperfection at the top boundary of the mesh with a wavenumber $k_{cr}$ and an amplitude $h=10^{-4} H$ as done in \cite{ciarletta2014pattern, ciarletta2016morphology}.
	
	The solution is found by using an incremental iterative Newton--Raphson method increasing (or decreasing in the fluid analogue case) the control parameter 
$\gamma$. In each iteration, the calculation is performed by using the linear algebra back-end PETSc (Portable, Extensible Toolkit for Scientific Computation) and the linear system is solved through a LU (Lower-Upper) decomposition. The code automatically adjust the increment of the control parameter if $\gamma$ is near the critical value $\gamma_{cr}$ or when the Newton--Raphson method fails to converge.
	
	Since secondary bifurcations may appear in such a dispersive problem, due to subharmonic resonance phenomena  in the
fully nonlinear regime, we performed further simulations
using the approach proposed in \cite{budday2015period}. Accordingly, we looked for period-doubling and period-tripling secondary bifurcations by using as computational domain the
sets $[0,\,2 m \pi/k_{cr}]\times [0,\,1]$ with $m = 2$ and $m=3$, respectively. However, we did not find any further
bifurcation in the parameters' range considered in the manuscript, in agreement with the experimental observations performed in the single layer case \cite{mora2014gravity}.
	\subsection{Numerical results}
	In the following, we report the results of the numerical simulation for the two cases under considerations.
	
	\subsubsection{Case (a): free surface instability ($\gamma>0$)}
		\begin{figure}[!ht]
	\centering
	\subfloat[$\alpha_\mu=0.5$]{\includegraphics[width=0.4\textwidth]{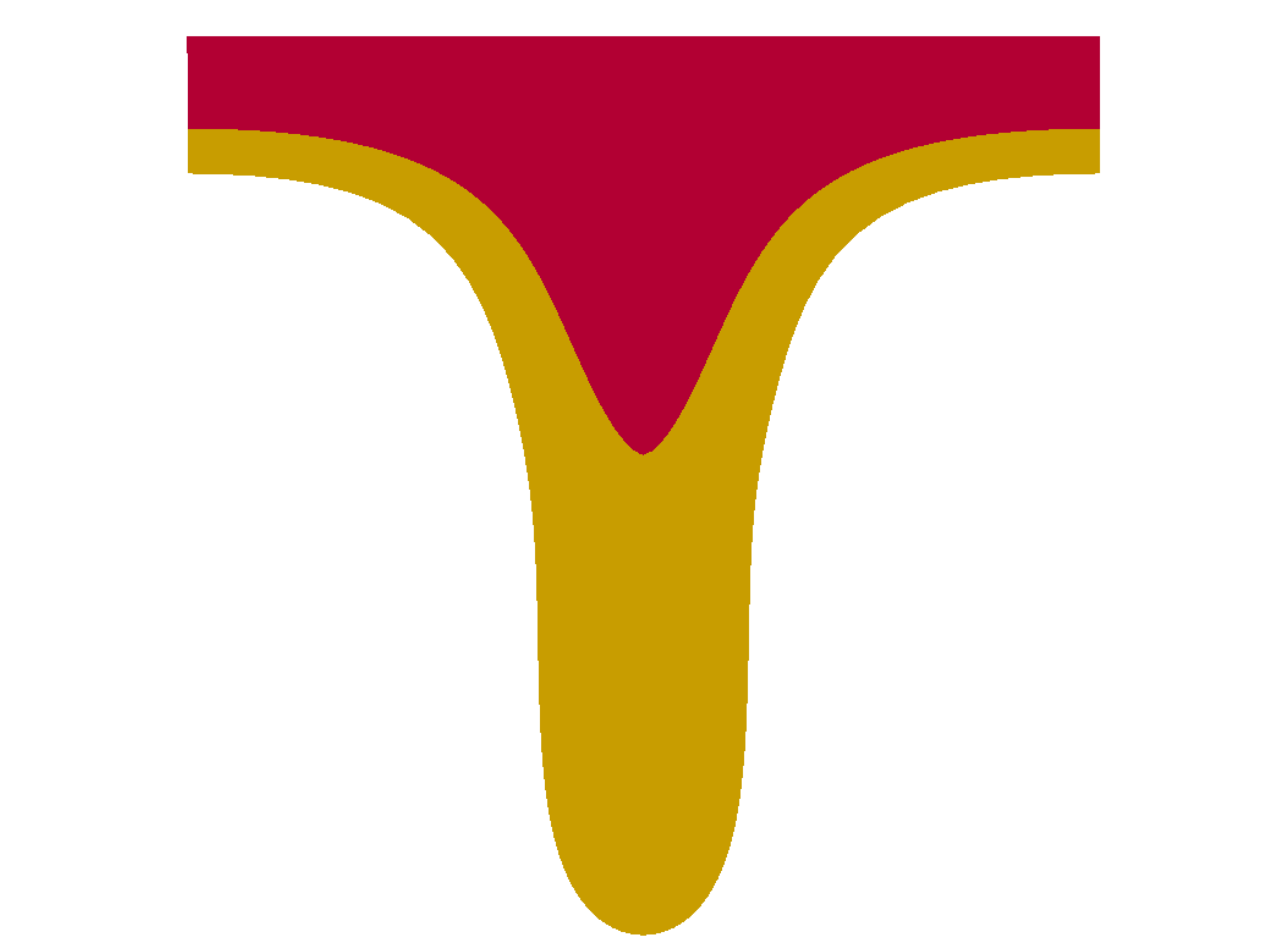}}
	\subfloat[$\alpha_\mu=2$]{\includegraphics[width=0.4\textwidth]{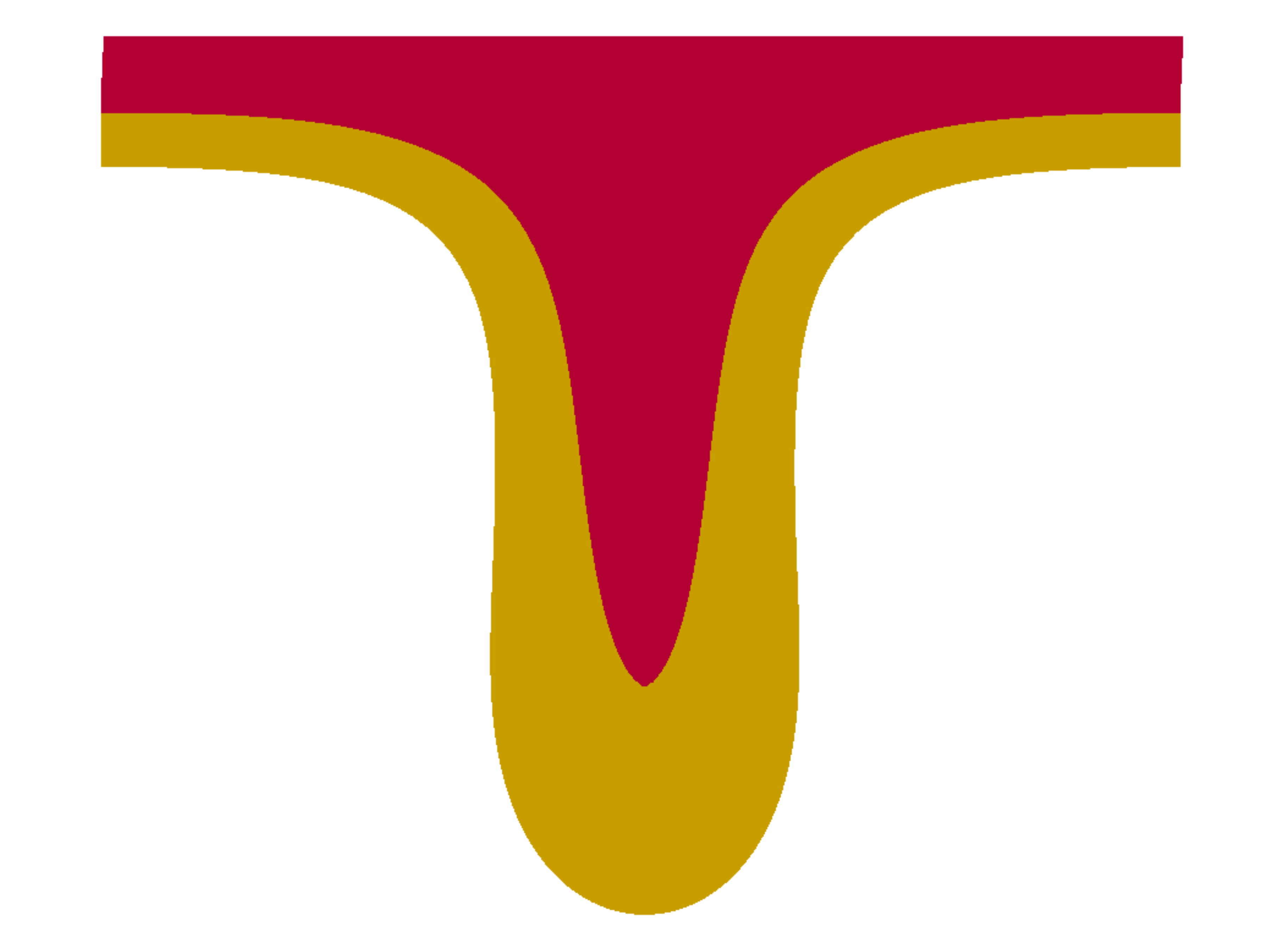}}\\
	\subfloat[$\alpha_\mu=0.5$]{\includegraphics[width=0.4\textwidth]{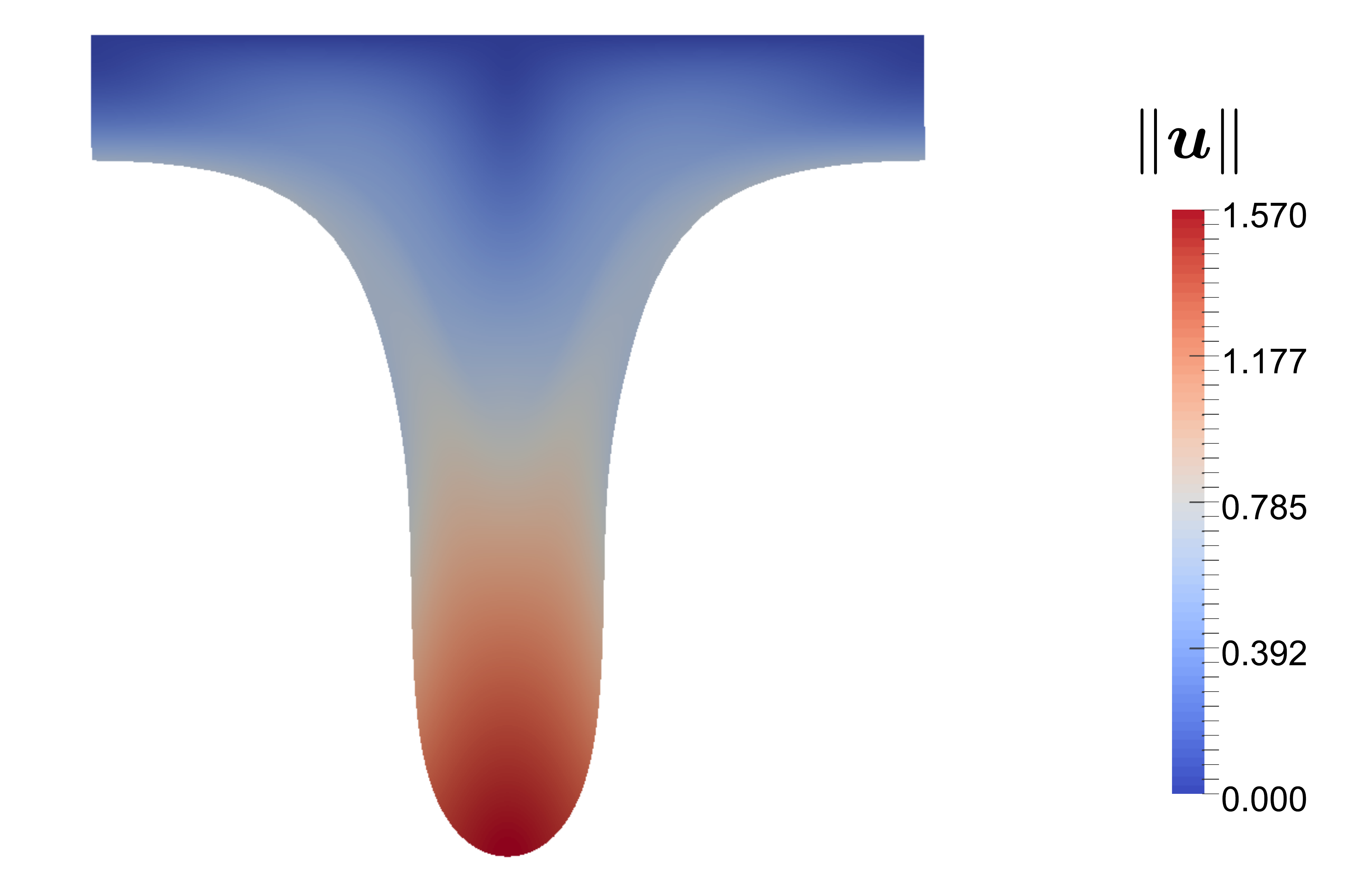}}
	\subfloat[$\alpha_\mu=2$]{\includegraphics[width=0.4\textwidth]{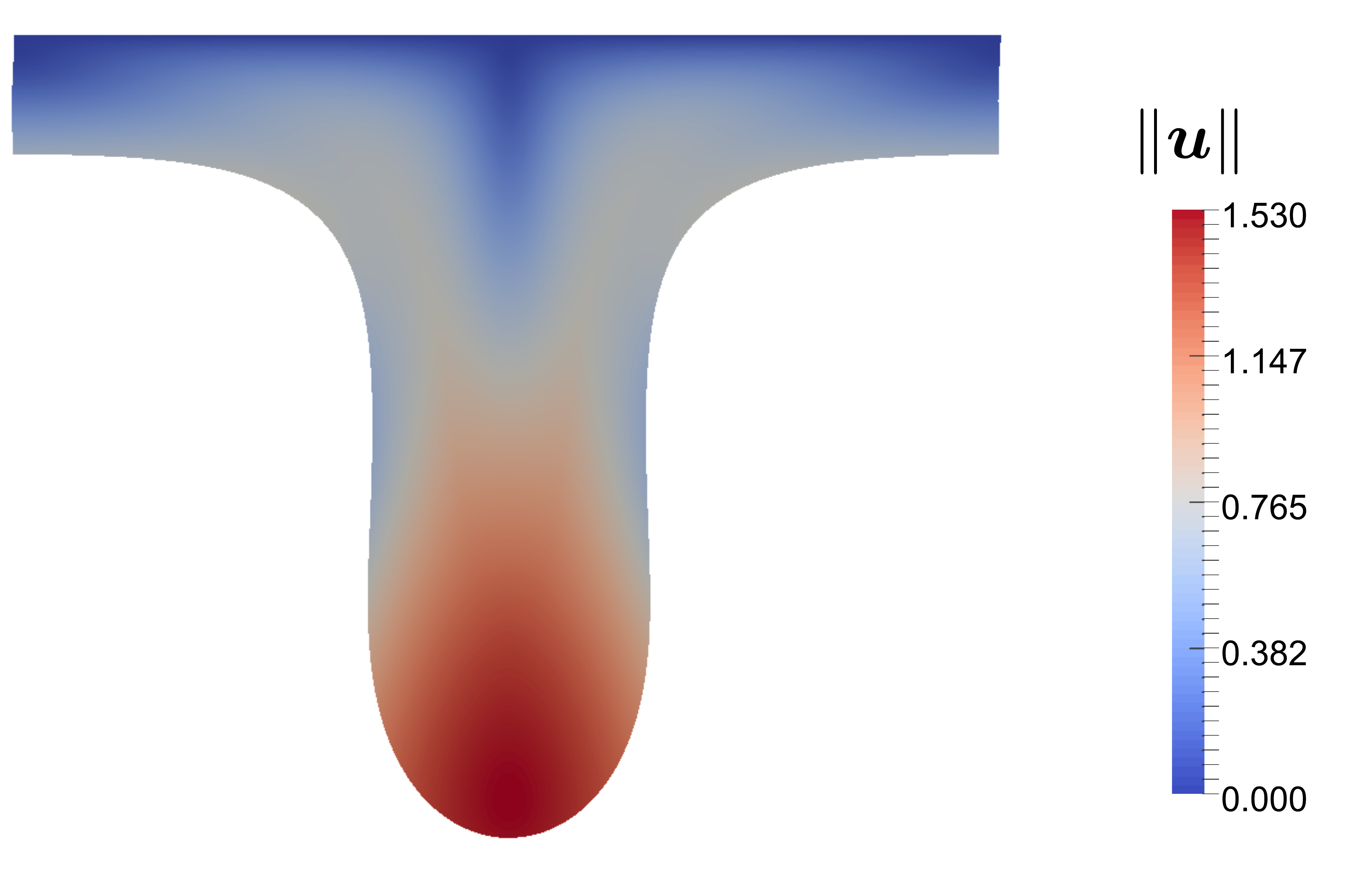}}\\
	\caption{Resulting fingering morphology  and displacement fields setting $\alpha_H=1$, $\alpha_\rho=1$ and (a, c) $\alpha_\mu=0.5$ and $\gamma=3.14$; (b, d) $\alpha_\mu=2$ and $\gamma=5.5$. In (c, d) the colorbars indicate the norm of the displacement.}
	\label{fig:shapegmag}
	\end{figure}	
\begin{figure}[!ht]
	\centering
	\subfloat[Plot of $\Delta h$ vs. $\gamma$]{\label{fig:dhnumgmag}\includegraphics[width=0.5\textwidth]{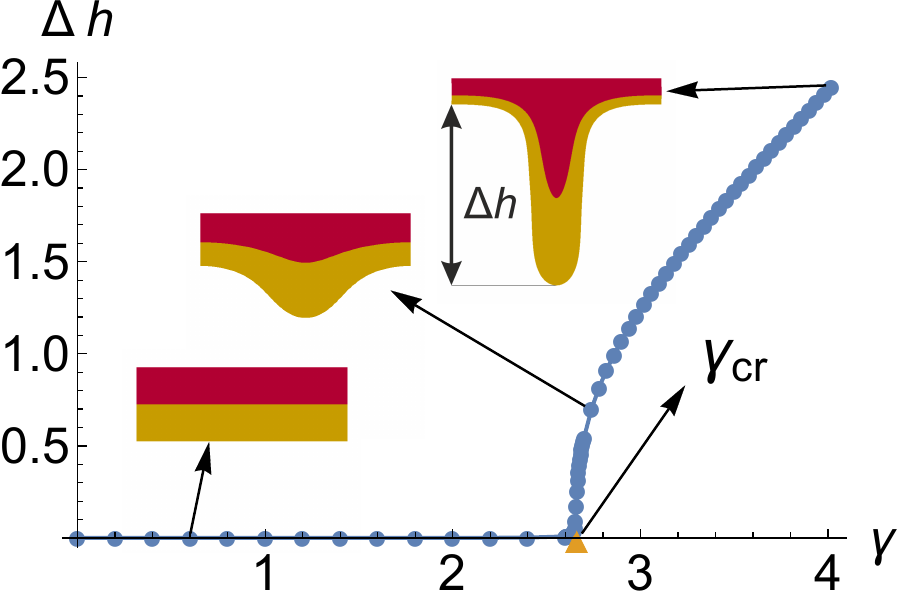}}
	\subfloat[Plot of $\Delta l/\lambda$ vs. $\gamma-\gamma_{cr}$]{\label{fig:dlnumgmag}\includegraphics[width=0.5\textwidth]{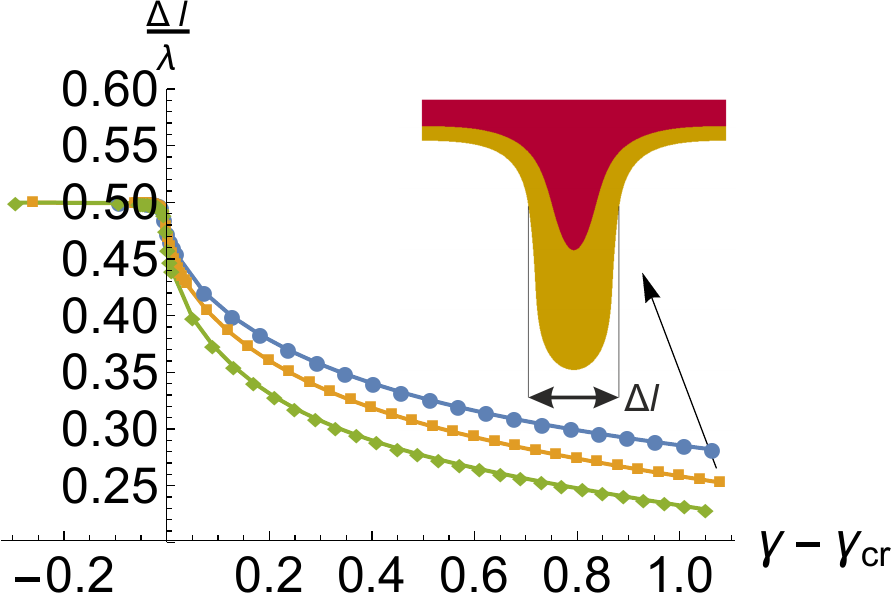}}
	\caption{(a) Numerical results showing the height of the fingers $\Delta h$ versus the order parameter $\gamma$, setting $\alpha_\rho=1$, $\alpha_H=1$, $\alpha_\mu=0.75$. The simulations validate the marginal stability threshold $\gamma_{cr}\simeq 2.66$ predicted by the linear stability analysis. (b) Plot of the normalized fingers' thickness $\Delta l/\lambda$ for $\alpha_\mu=0.5$ (green), $0.75$ (orange) and $2$ (blue).}
	\label{fig:numgmag}
	\end{figure}

	We first implement the case described in Figure~\ref{fig:dominio1},  setting $\alpha_\rho=1$ in order to mimic the behaviour of a slab made of two hydrogel layers. In Figure~\ref{fig:shapegmag}, we depict the results of the numerical simulations for two different values of $\alpha_\mu$.  In particular, we highlight that the deformation is localised at the free boundary of the body, and it evolves towards the formation of stable hanging digitations.  Let  $\Delta h$ be the maximum vertical distance of the points on the free surface whilst  $\Delta l$ be the horizontal distance between the points which have initial coordinates $(\lambda/4,\,H)$ and $(3/4\,\lambda,\,H)$, so that $\Delta l/\lambda=0.5$ if $\gamma<\gamma_{cr}$. Thus, we employ $\Delta h$ and $\Delta l$ to study the nonlinear evolution of the fingers' morphology.	
	
	As shown in Figure~\ref{fig:dhnumgmag}, we find that  the fingering height $\Delta h$ continuously increases as the control parameter $\gamma$ goes beyond its critical value. When performing a cyclic variation of the order parameter, where we first incremented $\gamma$ until a value $\gamma_{\text{max}}>\gamma_{cr}$ and later decreased it back to the initial value, we found that both $\Delta h$ and $\Delta l$ did not encounter any discontinuity, always following the same curve in both directions. Moreover, in the weakly nonlinear regime $\Delta h$ increases as the square root of the distance to threshold of the order parameter,  thus highlighting the presence of a supercritical pitchfork bifurcation.
	
	The shape and the thickness of these fingers strongly depend on the stiffness and  the thickness of the two layers. As shown in Figure~\ref{fig:dlnumgmag}, the fingers become thicker as we increase $\alpha_\mu$.
	
	We remark that the maximum diameter $h$ of the mesh elements is chosen as the maximum  value such that the resulting $\Delta h$ and $\Delta l/\lambda$  differ by less than $10^{-3}$ from the corresponding values obtained using a refined mesh with $h/2$.
	
	
%
%
%
	\subsubsection{Case (b): interfacial instability ($\gamma<0$)}

\begin{figure}[!b]
	\centering
	\subfloat[$\alpha_H=1$]{\includegraphics[width=0.4\textwidth]{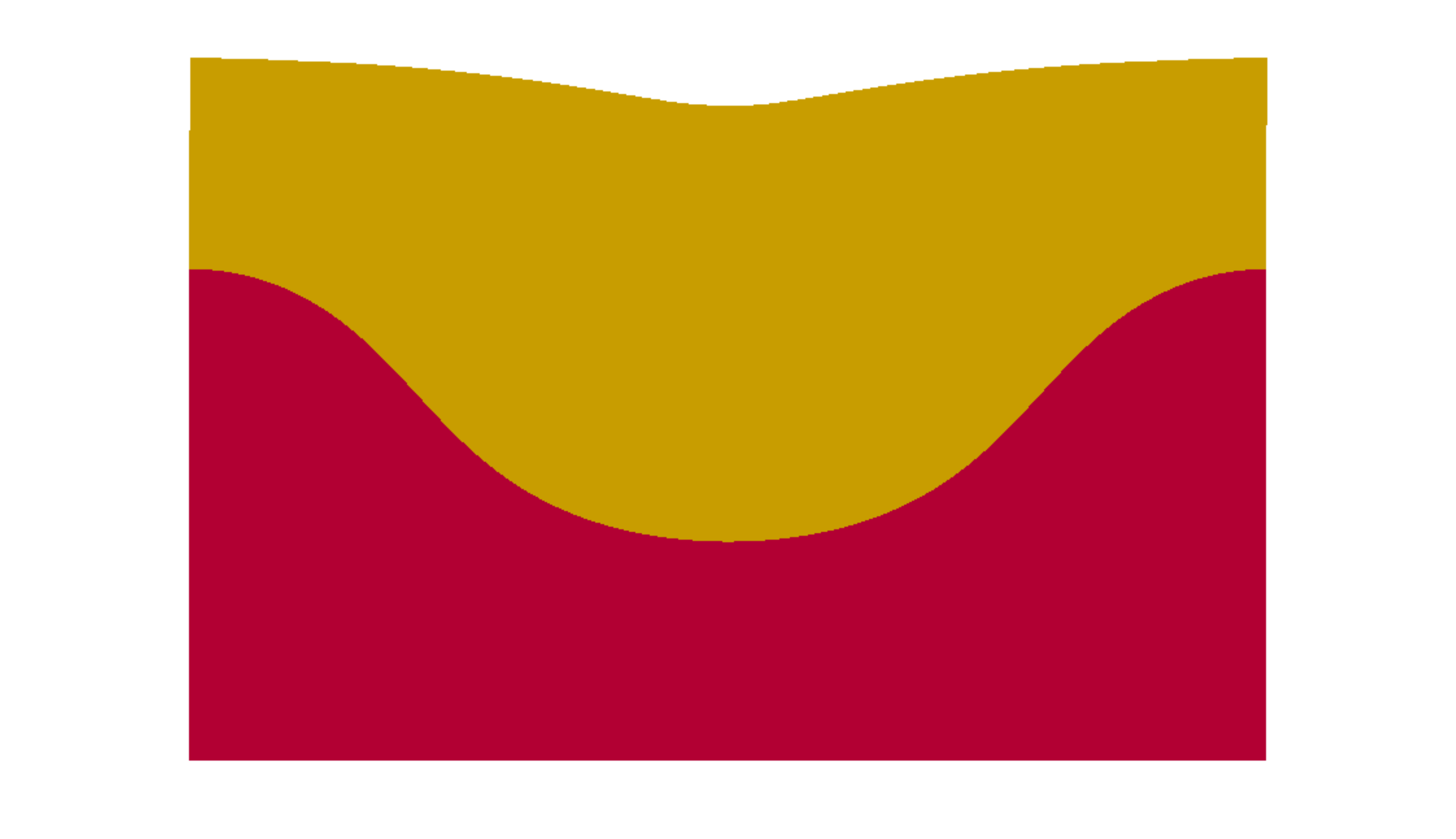}}
	\subfloat[$\alpha_H=0.25$]{\includegraphics[width=0.4\textwidth]{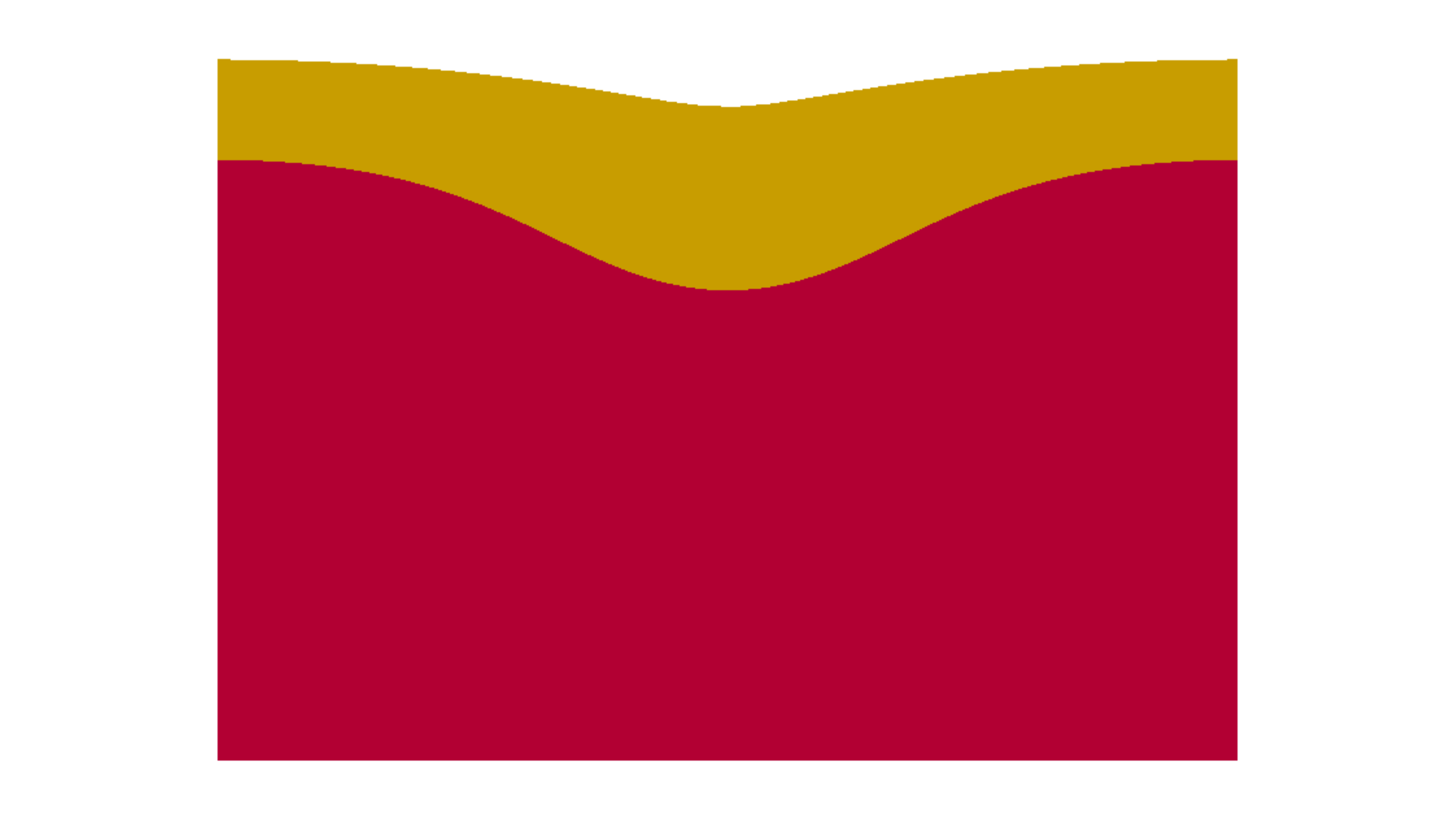}}\\
	\subfloat[$\alpha_H=1$]{\includegraphics[width=0.4\textwidth]{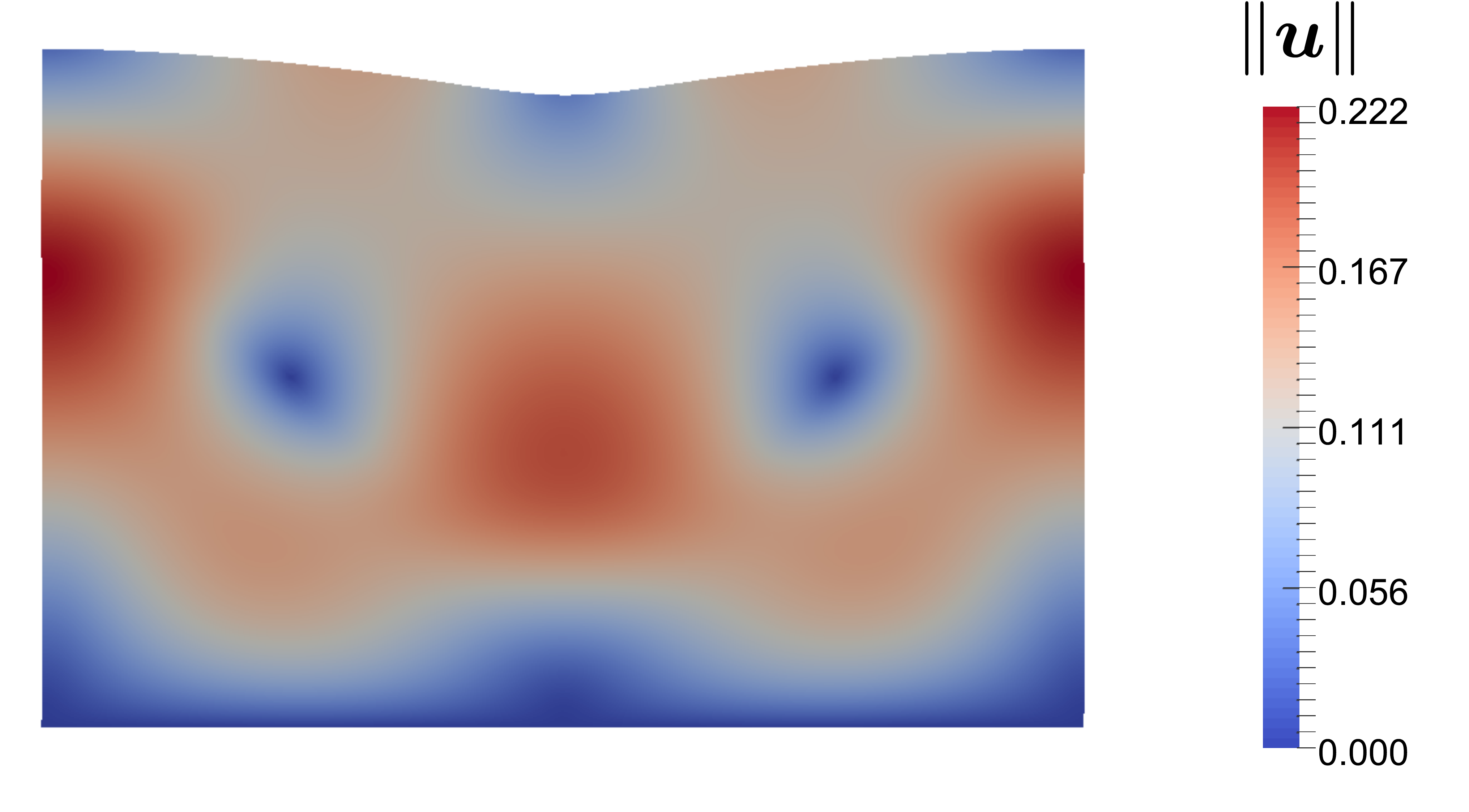}}
	\subfloat[$\alpha_H=0.25$]{\includegraphics[width=0.4\textwidth]{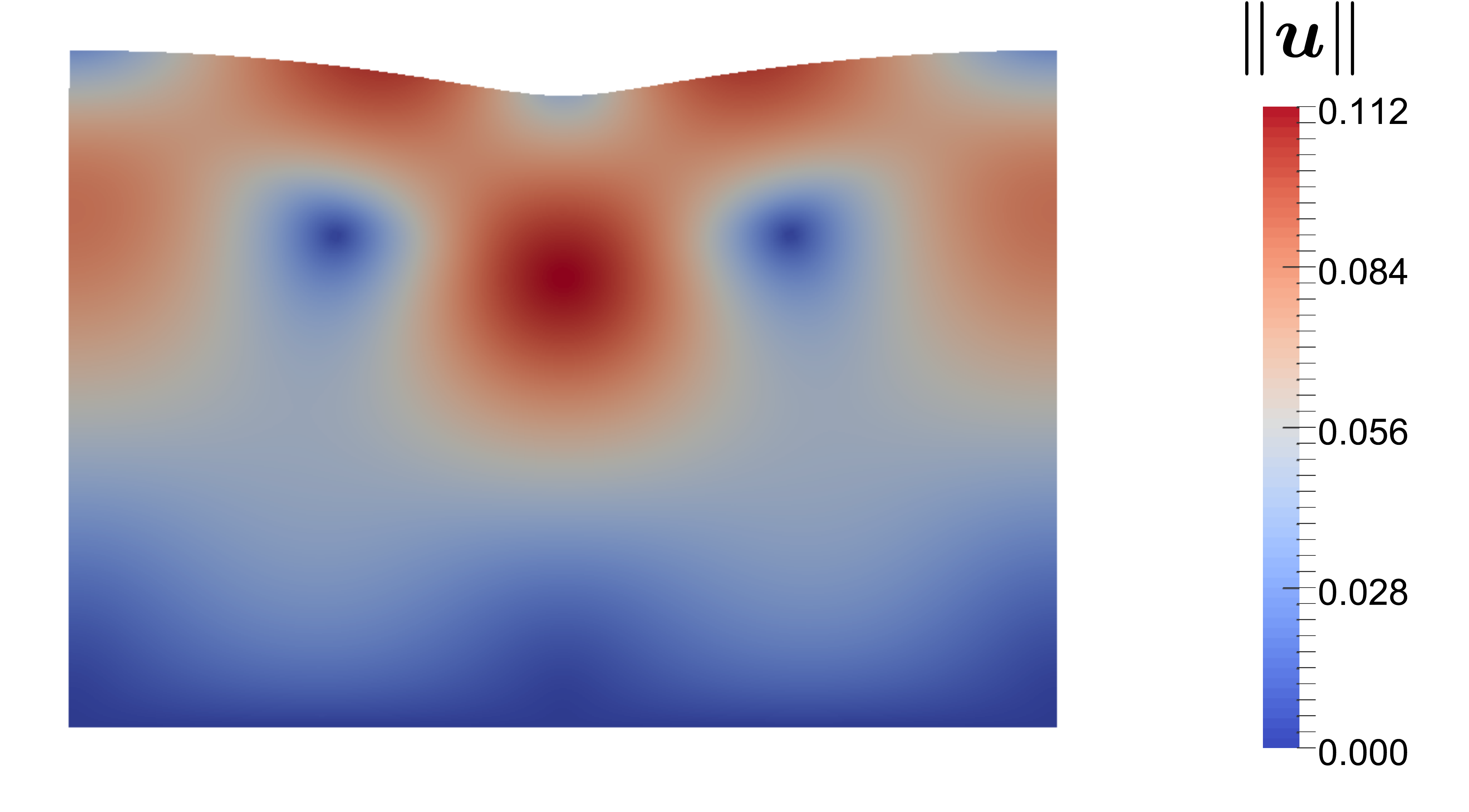}}
	\caption{Resulting morphology and displacement fields setting $\alpha_\mu=1$, $\alpha_\rho=2$ and (a, c) $\alpha_H=1$ and $\gamma=-12.36$, (b, d) $\alpha_H=0.25$ and $\gamma=-21.12$. In (c, d) the colorbars indicate the norm of the displacement.}
	\label{fig:shapegmin}
	\end{figure}	
	\begin{figure}[!t]
	\centering
	\subfloat[Plot of $\Delta h$ vs. $\gamma$]{\label{fig:fluid_dh}\includegraphics[width=0.5\textwidth]{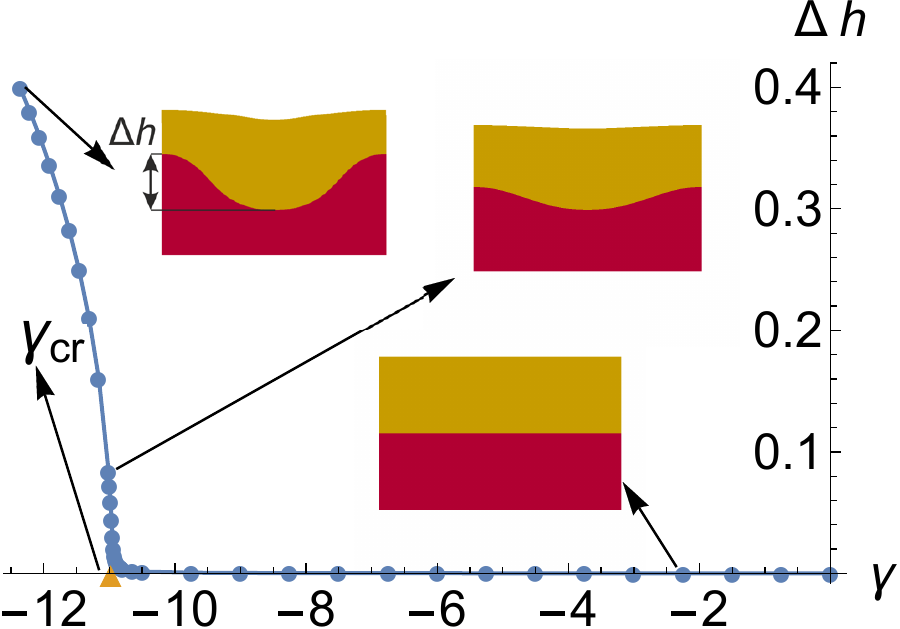}}
	\subfloat[Plot of $\Delta l/\lambda$ vs. $\gamma-\gamma_{cr}$]{\label{fig:fluid_dl}\includegraphics[width=0.5\textwidth]{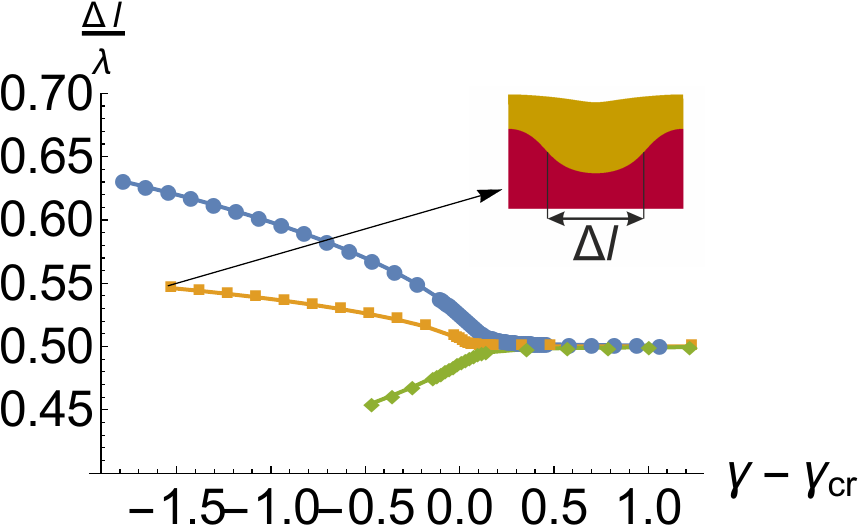}}
	\caption{(a) Numerical results showing the height $\Delta h$ of the undulation versus the order parameter $\gamma$, setting $\alpha_\rho=2$, $\alpha_H=1$, $\alpha_\mu=1$. The simulations validate the marginal stability threshold $\gamma_{cr}\simeq -10.97$ predicted by the linear stability analysis. (b) Plot of the normalized undulation thickness $\Delta l/\lambda$ for $\alpha_H=0.25$ (green), $1$ (orange) and $4$ (blue).}
	\label{fig:dhdl}
	\end{figure}

We now focus on the elastic analogue of the Rayleigh Taylor instability, occurring in the configuration depicted in Figure~\ref{fig:dominio2}. Here we set $\alpha_\rho=2$, so the top layer is heavier than the bottom one.
	
	The  simulation results are depicted in Figure~\ref{fig:shapegmin} for two different values of $\alpha_H$. In particular, we find a behavior similar to the Rayleigh--Taylor instability in fluids: the displacement is concentrated at the interface of the two layers forming a marginally stable undulation.
	
	Let $\Delta h$ denote here the maximum vertical distance of the points on the interface between the two layers whilst let $\Delta l$ be the horizontal distance between the points which have initial coordinates $(\lambda/4,\,H_a)$ and $(3/4\,\lambda,\,H_a)$, so that $\Delta l/\lambda=0.5$ if $\gamma>\gamma_{cr}$. In Figure~\ref{fig:dhdl}, we show the nonlinear evolution of such morphological parameters as a function of $\gamma$. As in the previous case, we measured the quantity $\Delta h$ decreasing the parameter $\gamma$ finding a continuous increase of the height of the undulation, as reported in Figure~\ref{fig:fluid_dh}. We highlight that the normalized thickness $\Delta l/\lambda$ strongly depends on the parameter $\alpha_H$, as we can see from Figure~\ref{fig:fluid_dl}.
	
	In fact, for thin soft layers the undulation decreases its width whilst decreasing $\gamma$ beyond its critical value, thus forming a digitation. Conversely,  the undulation width increases for thick top layers, thus forming a stable wrinkle.  In summary, $\Delta l/\lambda$ increases if the top layer is sufficiently thin, whilst it decreases if the top layer is above a critical thickness.
In both cases, the resulting morphology is perfectly reversible after cyclic variations of the order parameters, highlighting the presence of a supercritical pitchfork bifurcation.	

	\section{Discussion and concluding remarks}
	
	In this work, we used theoretical and computational tools to investigate the stability of a soft elastic bilayer subjected only to the bulk gravity force.
	
	Assuming that both layers are made of incompressible neo-Hookean materials, we have first formulated the boundary value problem in nonlinear elasticity considering that the slab attached on one end to a rigid substrate and it is traction-free at the other end. Considering the two configurations depicted in Figure~\ref{fig:dominio}, we have identified their basic undeformed solutions in Eq.~\eqref{eqn:basesol}, characterized by an hydrostatic pressure linearly varying on the thickness direction.
	
	Secondly, we have studied the linear stability by the means of the method of incremental deformations superposed on the basic elastic solution. We found that both configurations can undergo a morphological transition governed by the order parameter $\gamma$, representing the ratio between potential and elastic energies of the top layer. In particular, its critical value depends on three dimensionless parameters: $\alpha_H$, $\alpha_\mu$, $\alpha_\rho$ representing the thickness, shear moduli and density ratios between the layers, respectively.
	
	Thirdly, we have implemented a finite element code to solve the boundary value problem in the fully nonlinear instability regime. Other than validating the predictions of the linear stability analysis, the simulations have highlighted the nonlinear evolution of the characteristic patterns.

	Compared to the classic Rayleigh-Taylor hydrodynamic instability, not surprisingly we have found that elastic effects tend to stabilize the dynamics of the surface undulations forming beyond the linear stability threshold. Nonetheless, we obtained a rich morphological diagram with respect to both geometric and elastic parameters. In the following, we briefly discuss the main results for the two cases under consideration.
	
	If the body hangs below a rigid wall, as depicted in Figure~\ref{fig:dominio1}, we find that there always exists a critical value for the order parameter, driving a morphological transition localised at the free surface. Such an shape instability is favoured if the bottom layer is softer and thicker than the top one, having a critical horizontal wavelength of the same order as the body thickness. In the nonlinear regime, this critical undulation evolves towards forming a digitation, whose characteristic penetration length continuously increases beyond the linear stability threshold, highlighting the existence of a supercritical pitchfork bifurcation.
	
	If the body is attached to a rigid substrate at the bottom surface, as depicted in Figure~\ref{fig:dominio2}, a morphological transition can occur if and only if the top layer has a higher density than the bottom one. Similarly to the previous case, the onset of an elastic bifurcation is favoured by a softer and thicker bottom layer compared to the top one, with a critical wavelength of the same order as the body thickness. However, an important difference is that the shape instability is localised at the interface between the two layers, displaying two characteristic nonlinear patterns. If the top layer is thinner than the bottom one, the undulation evolves towards forming finger-like protrusions, whilst in the opposite geometrical limit a stable wrinkling occurs.

	In summary, we have characterized the shape instabilities occurring in a soft elastic bilayer subjected only to the action of the gravity bulk force. Unlike the Rayleigh-Taylor instabilities in fluids, we have demonstrated that the nonlinear elastic effects saturate the dynamic instability of the bifurcated solutions, displaying a rich morphological diagram where both digitations and stable wrinkling can emerge. The results of this work provide important guidelines for the design of novel soft systems with tunable shapes. In fact, the possibility to control by external stimuli both the geometric and the elastic properties in smart materials, such as hydrogels or dielectric elastomers \cite{pelrine2001dielectric}, can be used to provoke morphological transitions on demand \cite{gladman2016biomimetic}. Morphological changes in such soft devices may be used, for example, to selectively change the surface roughness (e.g. to perform drag reduction in fluid-structure interactions \cite{dean2010shark}) or to fabricate tailor-made patterns (e.g. to design adaptive material scaffolds \cite{saha2010surface}).
	\appendix

	\section{Appendix -- Structure of the matrix $\tens{M}$}
	\label{sec:AA}
	
	We report the matrix $\tens{M}$ we used in the equation Eq.~\eqref{eqn:matrice}. We split it into 16 blocks:
	\[	
	\tens{M}=\begin{bmatrix}
	\tens{\bf{0}} 			&\tens{\bf{0}}		&\tens{M}_{13}	&\tens{M}_{14}\\
	\tens{M}_{21}		&\tens{M}_{22} 	&\tens{\bf{0}} 		&\tens{\bf{0}}\\
	\tens{M}_{31}		&\tens{M}_{32}	&\tens{M}_{33}	&\tens{M}_{34}\\
	\tens{M}_{41}		&\tens{M}_{42}	&\tens{M}_{43}	&\tens{M}_{44}\\
	\end{bmatrix},
	\]
	where $\bf{0}$ is the null $2\times 2$ matrix and:
	\begingroup
\everymath{\scriptstyle}
	\begin{gather*}
	\tens{M}_{13}=\begin{bmatrix}
	 \tilde{k} & \alpha_H \tilde{k}+\tilde{k}-1 \\
 -2 \tilde{k} \alpha_\mu-\alpha_\rho \gamma  & -(\alpha_H+1) (2 \tilde{k} \alpha_\mu+\alpha_\rho \gamma ) \\
	\end{bmatrix},\\
	\tens{M}_{14}=\begin{bmatrix}
e^{2 \alpha_H \tilde{k}} \tilde{k} & e^{2 \alpha_H \tilde{k}} (\alpha_H \tilde{k}+\tilde{k}+1) \\
 e^{2 \alpha_H \tilde{k}} (2 \tilde{k} \alpha_\mu-\alpha_\rho \gamma ) & e^{2 \alpha_H \tilde{k}} (\alpha_H+1) (2 \tilde{k} \alpha_\mu-\alpha_\rho \gamma ) \\
	\end{bmatrix},\quad
	\tens{M}_{21}=\begin{bmatrix}
1 & 0 \\
 -\tilde{k} & 1 \\
	\end{bmatrix},\\
	\tens{M}_{22}=\begin{bmatrix}
 1 & 0 \\
 \tilde{k} & 1 \\
	\end{bmatrix},\quad	
	\tens{M}_{31}=\begin{bmatrix}
1 & 1 \\
 -\tilde{k} & 1-\tilde{k} \\
	\end{bmatrix},\quad
	\tens{M}_{32}=\begin{bmatrix}
 e^{2 \tilde{k}} & e^{2 \tilde{k}} \\
 e^{2 \tilde{k}} \tilde{k} & e^{2 \tilde{k}} (\tilde{k}+1) \\
	\end{bmatrix},\\
	\tens{M}_{33}=\begin{bmatrix}
 -1 & -1 \\
 \tilde{k} & \tilde{k}-1 \\
	\end{bmatrix},\quad
	\tens{M}_{34}=\begin{bmatrix}
 -1 & -1 \\
 -\tilde{k} & -\tilde{k}-1 \\
	\end{bmatrix},\\
	\tens{M}_{41}=\begin{bmatrix}
 -\tilde{k} (\alpha_H \alpha_\rho \gamma -2) & \tilde{k} (2-\alpha_H \alpha_\rho \gamma )-2 \\
 \tilde{k} (\alpha_H \alpha_\rho \gamma -2)-\gamma  & \tilde{k} (\alpha_H \alpha_\rho \gamma -2)-(\alpha_H \alpha_\rho+1) \gamma  \\
	\end{bmatrix},\\
	\tens{M}_{42}=\begin{bmatrix}
-e^{2 \tilde{k}} \tilde{k} (\alpha_H \alpha_\rho \gamma -2) & e^{2 \tilde{k}} (\tilde{k} (2-\alpha_H \alpha_\rho \gamma )+2) \\
 -e^{2 \tilde{k}} (\gamma +\tilde{k} (\alpha_H \alpha_\rho \gamma -2)) & -e^{2 \tilde{k}} (\alpha_H \alpha_\rho \gamma +\gamma +\tilde{k}
   (\alpha_H \alpha_\rho \gamma -2)) \\
	\end{bmatrix},\\
\tens{M}_{43}=\begin{bmatrix}
 \tilde{k} (\alpha_H \alpha_\rho \gamma -2 \alpha_\mu) & \alpha_H \tilde{k} \alpha_\rho \gamma -2 (\tilde{k}-1) \alpha_\mu \\
 2 \tilde{k} \alpha_\mu-\alpha_H \tilde{k} \alpha_\rho \gamma +\alpha_\rho \gamma  & (\alpha_H+1) \alpha_\rho \gamma +\tilde{k} (2 \alpha_\mu-\alpha_H \alpha_\rho \gamma ) \\
	\end{bmatrix},\\
	\tens{M}_{44}=\begin{bmatrix}
  \tilde{k} (\alpha_H \alpha_\rho \gamma -2 \alpha_\mu) & \alpha_H \tilde{k} \alpha_\rho \gamma -2 (\tilde{k}+1) \alpha_\mu \\
 -2 \tilde{k} \alpha_\mu+\alpha_H \tilde{k} \alpha_\rho \gamma +\alpha_\rho \gamma  & (\tilde{k} \alpha_H+\alpha_H+1) \alpha_\rho \gamma -2 \tilde{k} \alpha_\mu \\
	\end{bmatrix}.
	\end{gather*}
	\endgroup

\section{Appendix -- Expressions of the coefficients $c_j$}
\label{sec:AB}

The coefficient $c_1$ of equation Eq.~\eqref{eqn:eqdi2grado} is given by
\begin{align*}
\scriptstyle c_1=&\scriptstyle 2 \tilde{k}^2 \alpha_\mu \left(-2 \left(\alpha_\mu^2-1\right) \left(2 \alpha_H^2 \tilde{k}^2+1\right) \cosh (2\tilde{k})+2 \left(4 \alpha_H^2 \tilde{k}^4 (\alpha_\mu-1)^2+\right.\right.\\
&\left.\scriptstyle +2 \tilde{k}^2 \left(\alpha_H^2 \left(\alpha_\mu^2+1\right)+4 \alpha_H
   \alpha_\mu+\alpha_\mu^2+1\right)-\left(2 \tilde{k}^2+1\right) \left(\alpha_\mu^2-1\right) \cosh (2 \alpha_H
   \tilde{k})+\alpha_\mu^2+1\right)+\\
   &\scriptstyle\left. +(\alpha_\mu-1)^2 \cosh (2 (\alpha_H-1) \tilde{k})+(\alpha_\mu+1)^2 \cosh (2 (\alpha_H+1)
   \tilde{k})\right),
\end{align*}
whereas $c_2$ is
\begin{align*}
\scriptstyle
c_2= &\scriptstyle \tilde{k} \left(4 \alpha_H^2 \tilde{k}^2 \alpha_\rho \alpha_\mu \sinh (2 \tilde{k})-4 \alpha_H^2 \tilde{k}^2 \alpha_\mu \sinh (2 \tilde{k})+4 \tilde{k} \left(\alpha_\mu
   \left(2 \alpha_H \tilde{k}^2 (\alpha_H-(\alpha_H+2) \alpha_\rho)+\alpha_\rho+1\right)+\right.\right.\\
   &\scriptstyle\left.+\alpha_H \left(2 \tilde{k}^2+1\right) \alpha_\rho+\alpha_\mu^2
   \left(2 \alpha_H \tilde{k}^2+\alpha_H\right)\right)
   -4 \tilde{k}^2 \alpha_\rho \sinh (2 \alpha_H \tilde{k})+4 \tilde{k}^2 \alpha_\mu^2 \sinh
   (2 \alpha_H \tilde{k})+\\
   &\scriptstyle +4 \alpha_H \tilde{k} \cosh (2 \tilde{k}) \left(\alpha_\rho-\alpha_\mu^2\right)-2 \alpha_\mu \cosh (2 \alpha_H \tilde{k}) (2 \tilde{k}
   (\alpha_\rho-1)+(\alpha_\rho+1) \sinh (2 \tilde{k}))-2 \alpha_\rho \sinh (2 \alpha_H \tilde{k})+\\
   &\scriptstyle-\alpha_\rho \sinh (2 (\alpha_H+1) \tilde{k}) +\alpha_\rho
   \sinh (2 \tilde{k}-2 \alpha_H \tilde{k})-4 \alpha_\mu^2 \sinh ^2(\tilde{k}) \sinh (2 \alpha_H \tilde{k})+\\
   &\scriptstyle\bigl.+4 \alpha_\rho \alpha_\mu \sinh (\tilde{k})
   \cosh (\tilde{k})-4 \alpha_\mu \sinh (\tilde{k}) \cosh (\tilde{k})\bigr),
\end{align*}
and the expression of $c_3$ is
\begin{align*}
\scriptstyle
c3=&\scriptstyle
-\frac{1}{2} (\alpha_\rho-1) \alpha_\rho \left(2 \left(2 \tilde{k}^2 (2 \alpha_H+\alpha_\mu)-\left(2 \tilde{k}^2+1\right) \alpha_\mu
   \cosh (2 \alpha_H \tilde{k})-2 \tilde{k} (\alpha_H \sinh (2 \tilde{k})+\right.\right.\\
   &\scriptstyle\left.\left.+\sinh (2 \alpha_H \tilde{k}))+\alpha_\mu\right)+(\alpha_\mu-1) \cosh
   (2 (\alpha_H-1) \tilde{k})+(\alpha_\mu+1) \cosh (2 (\alpha_H+1) \tilde{k})-2 \alpha_\mu \cosh (2 \tilde{k})\right).
\end{align*}
\enlargethispage{20pt}


\section*{Funding}This work has been partially supported by \emph{Progetto Giovani GNFM 2016}
funded by the National Group of Mathematical Physics (GNFM - INdAM) and by the MFAG AIRC Grant 17412.

\bibliographystyle{ieeetr}
\bibliography{refs}

\end{document}